\documentclass[amsmath, amssymb, twocolumn, aps, prx, superscriptaddress]{revtex4-1}

\usepackage{mathtools}
\usepackage{mathrsfs, amsmath, wasysym}
\usepackage{bbold}
\usepackage[mathscr]{eucal}
\usepackage{xfrac}
\usepackage{graphicx}
\usepackage{dcolumn}
\usepackage{bm}
\usepackage{color}
\usepackage{tabularx}
\usepackage{natbib}
\usepackage[colorlinks,linkcolor=blue,citecolor=blue,urlcolor=blue]{hyperref}
\usepackage[normalem]{ulem}
\usepackage[all]{hypcap}
\usepackage[dvipsnames,usenames,table]{xcolor}
\usepackage{braket,mleftright}

\newcommand{\nocontentsline}[3]{}
\newcommand{\tocless}[2]{\bgroup\let\addcontentsline=\nocontentsline#1{#2}\egroup}

\newcommand{\bs}{{\bf s}}
\newcommand{\bS}{{\bf S}}

\newcommand{\bT}{{\bf T}}

\newcommand{\bsigma}{\boldsymbol{\sigma} }

\newcommand{\be}{\begin{equation}}
\newcommand{\ee}{\end{equation}}
\newcommand{\beg}{\begin{gather}}
\newcommand{\eeg}{\end{gather}}
\newcommand{\beq}{\begin{eqnarray}}
\newcommand{\eeq}{\end{eqnarray}}
\newcommand{\bea}{\begin{align}}
\newcommand{\eea}{\end{align}}
\newcommand{\beqq}{\begin{eqnarray*}}
\newcommand{\eeqq}{\end{eqnarray*}}

\newcommand{\up}{\uparrow}
\newcommand{\down}{\downarrow}

\begin{document}

\title{Kondo-Heisenberg toy models: Comparison of exact results and spin wave expansion}

\author{M. Frakulla}
\thanks{These authors contributed equally to this work.}
\affiliation{Department of Physics, Drexel University, Philadelphia, PA 19104, USA}%

\author{J. Strockoz}
\thanks{These authors contributed equally to this work.}
\affiliation{Department of Physics, Drexel University, Philadelphia, PA 19104, USA}%

\author{D. S. Antonenko}
\affiliation{Department of Physics, Yale University, New Haven, CT 06520, USA}%

\author{J. W. F. Venderbos}
\affiliation{Department of Physics, Drexel University, Philadelphia, PA 19104, USA}%
\affiliation{Department of Materials Science \& Engineering, Drexel University, Philadelphia, PA 19104, USA}%

\begin{abstract}
In this paper we study a class of exactly solvable Kondo-Heisenberg toy models in one dimension, with the goal of comparing the exact low-energy excitations of the ferromagnetic ground state to the approximate solution obtained from spin wave theory. In doing so we employ a recently introduced strong coupling $1/S$ spin wave expansion, which effectively describes excitations of the total spin $S\pm 1/2$ on a given site (i.e., sum of local moment and electron spin). We further make use of the fact that the ground state of Kondo lattice models with quantum spins and a single electron is a ferromagnet, and that the magnetic excitations of the ferromagnet can be exactly determined. We demonstrate that the energies and eigenstates of the spin waves are in full agreement with the exact solution order-by-order in $1/S$ and $t/J_K$, the strong coupling expansion parameter. In the specific case of antiferromagnetic Kondo coupling, when the exact ground state wave function describes spin polaron, we show that the electron operators of the spin wave formalism precisely correspond to the spin polaron states. More broadly, the study of Kondo-Heisenberg toy models is shown to provide insight into the fundamental distinction between itinerant Kondo magnets and Heisenberg magnets. 
\end{abstract}

\date{\today}


\maketitle



\section{Introduction \label{sec:intro}}

The Kondo lattice model provides a simple and compelling framework for studying a variety of phenomena observed in strongly correlated electron systems. Magnetism arising from an interaction between itinerant charge carriers and a lattice of local moment spins is a prominent example of such phenomena. When studying magnetic ordering phenomena in the context of the Kondo lattice model, it is natural, and in fact common, to treat the local moments as classical spins~\cite{Izyumov:2001p109,Dagotto:2001p1,Batista:2016p084504}. This is appropriate for determining the classical magnetic ground state, but does not give access to low-energy magnetic excitations. A standard approach to addressing the properties of magnetic excitations is to perform a systematic $1/S$ spin wave expansion around the classical ground state~\cite{Holstein:1940p1098,Holstein:1941p388,Auerbach1994}. In itinerant magnets described by the Kondo lattice model this is more involved than in magnetic insulators described by Heisenberg models, since in the former the spin wave bosons are coupled to the itinerant charge carriers~\cite{Kubo:1972p21,Furukawa:1996p1174,Wang:1998p7427}. 

In previous work we have introduced a spin wave expansion for itinerant Kondo lattice magnets in the regime of large Kondo coupling $J_K$~\cite{our_paper}. This strong coupling $1/S$ expansion is based on a canonical Schrieffer-Wolff transformation performed after bosonizing the local moment spins via the Holstein-Primakoff (HP) substitution. The canonical transformation is designed to successively remove electron spin-flip terms, since these change the number of low- and high-energy electrons in the strong coupling regime. Due its reliance on a canonical transformation we refer to the resulting spin wave expansion as the canonical spin wave expansion, which has a number of appealing properties. First, it is formulated in terms of the natural degrees of freedom of the problem in the strong coupling regime: electrons in a state of total spin $S\pm 1/2 $ at each site (i.e., eigenstates of the quantum Kondo coupling), and bosons which correspond to spin wave excitations of the total spin (i.e., the total spin of electron and local moment). Second, projecting out the high-energy electrons (determined by the sign of $J_K$) yields a Hamiltonian for the low-energy dynamics of effectively spinless fermions interacting with spin wave bosons. This effective Hamiltonian accounts for the coupling to the high energy states perturbatively in $t/J_K$, where $t$ is the hopping. A third appealing property are the straightforward extensions and generalizations to systems that require a more involved description, for instance due to the importance of spin-orbit coupling, or the interplay between magnetism and superconductivity.


The purpose of this paper is to apply the canonical spin wave expansion to a class of Kondo-Heisenberg (KH) toy models for which exact solutions are available. The availability of exact solutions allows for a detailed comparison between the exact spectrum of the toy model Hamiltonians and the approximate solution obtained from the spin wave expansion scheme. More specifically, when the exact ground state of such KH toy models is a ferromagnet, i.e., a state with maximal total spin $T$, the energies and wave functions of the exact low-energy excitations can be compared to the energies and wave functions of the spin wave excitations. Such comparison is a key diagnostic for the adequacy and accuracy of the spin wave expansion. 

\footnotetext[1]{For a discussion of the relation between the canonical spin wave expansion and the spin wave expansion proposed in Ref.~\onlinecite{Shannon:2002p104418}, see Ref.~\onlinecite{our_paper}.}

This approach follows an earlier study by Shannon~\cite{Shannon:2001p6371}, in which the exact solution of a two-site Kondo dimer problem is compared to the predictions of a strong coupling spin wave expansion proposed by Shannon and Chubukov~\cite{Shannon:2002p104418,Note1}. Shannon demonstrated that the exact solution of the two-site problem is indeed correctly reproduced by the spin wave expansion order-by-order in $1/S$. This is a nontrivial result in the context of an itinerant Kondo model, since the classical limit of the on-site Kondo coupling, in which the spins are treated classically, is manifestly different from the quantum Kondo coupling. Therefore, order-by-order agreement with all characteristics of the exact excitations of an itinerant Kondo ferromagnet shows that the spin wave expansion correctly captures the quantum nature of the local moment spins. 

In this work we apply a similar approach to KH toy models, which include a Heisenberg exchange coupling $J$ between the local moments, and similarly demonstrate that the canonical spin wave expansion introduced in Ref.~\onlinecite{our_paper} correctly reproduces the exact solutions order-by-order in $1/S$, as well as in the perturbative expansion parameters $t/J_K$ and $J/J_K$. We first consider a KH model of two spins and one electron, which we refer to as the KH dimer. We compute and discuss the full exact spectrum and determine when the ferromagnetic ground state becomes unstable as the Heisenberg coupling $J$ is increased. The exact results are then compared to the predictions from spin wave theory. 

The second model we consider is the one-dimensional (1D) KH chain in the extremely dilute limit, i.e., the limit of a single electron. The one-electron Kondo model is known to admit a number of rigorous results regarding the ground state and the excitation spectrum, and is therefore a useful toy model~\cite{Ueda:1991p167,Sigrist:1991p2211,Sigrist:1992p13838,Tsunetsugu:1997p809}. In particular, it has been proven that the ground state is a ferromagnet, such that a comparison between the exact excitation spectrum and spin wave theory is meaningful. Here we revisit the 1D Kondo chain with $N$ sites, generalize it to arbitrary spin $S$~\cite{Henning:2012p085101,Masui:2022p014411}, and include an antiferromagnetic nearest neighbor coupling $J$ between the local moments. It is worth pointing out that the KH dimer may be viewed as the $N=2$ limit of the general 1D model, but we nonetheless consider it separately since the dimer is analytically solvable for all values of total spin $T$, whereas the $N$-site chain is solvable only in the invariant subspaces of total spin $T = NS\pm 1/2$. This is sufficient for our purposes, since it allows us to study both the structure of the ferromagnetic ground state, as well as its lowest energy magnetic excitations. In particular, when the Kondo exchange coupling is ferromagnetic, the ground state has total spin $T = NS+1/2$ and the magnetic excitations are located in the total spin $T = NS-1/2$ sector. We present a detailed analysis of the magnetic excitations and show that the canonical spin wave expansion captures the exact solution in the $T = NS-  1/2$ subspace order-by-order in the expansion parameters.

When the Kondo coupling is antiferromagnetic the ground state has total spin $T = NS-1/2$ and the electron forms a bound state with local moment magnons~\cite{Ueda:1991p167,Tsunetsugu:1997p809}. Such a bound state corresponds to a spin polaron and we address the structure of this spin polaron in detail. We will demonstrate that in the case of (strong) antiferromagnetic Kondo coupling the electron degrees of freedom of the canonical spin wave expansion precisely correspond to such spin polaron states. In particular, the spin wave expansion scheme both correctly captures the binding energy of the spin polaron and its wave function. 

On a more general and fundamental level, the KH toy models considered here provide important insight into the nature of itinerant Kondo lattice ferromagnets and the way in which these are fundamentally different from Heisenberg ferromagnets. In a Heisenberg ferromagnet, the spin wave excitations are exact eigenstates of the Hamiltonian, such that there is perfect agreement between the exact spectrum and linear spin wave theory. The KH toy models clearly demonstrate that this is very different in Kondo lattice ferromagnets, in which the spin wave excitations are manifestly not exact eigenstates of the Hamiltonian.  As we show, the exact energies of the elementary magnetic excitations can only be matched by spin wave theory order-by-order in $1/S$. This implies that quantum corrections beyond linear spin wave theory are important in itinerant magnets. Our study therefore offers a deeper understanding of the quantum structure of itinerant Kondo lattice magnets, in particular ferromagnets, and to what extent a spin wave theory can capture this.

\section{Kondo-Heisenberg dimer \label{sec:kondo-exact}}

We begin by considering the KH dimer, which consists of just two sites (labeled $i=1,2$) and is defined by the Hamiltonian
\be
H =  -t (c^\dagger_1 c_2+c^\dagger_2 c_1)-\frac{J_K}{S}\sum_{i=1}^2 \bS_i \cdot \bs_i + \frac{J}{S^2}\bS_1 \cdot \bS_2, \label{eq:H_KH}
\ee
where $t $ is the hopping, $J_K$ is the on-site Kondo coupling, and $J$ is the exchange coupling between the spins. The electron operators are $c_{i\sigma}$ with $\sigma=\up,\down$, and $\bs_i = c^\dagger_i \bsigma c_i/2$. Note that here we have defined the coupling constants such that in the classical limit, when the spins are treated as classical vectors of length $S$, the splitting between the locally aligned and anti-aligned electrons is $J_K$, and the exchange energy is $J \cos \theta$ (with $\theta$ the angle between spins). 

In general, the physical phenomena described by the Kondo lattice model are rather sensitive to the electron density. Here, in this toy model problem, we fix the ``electron density'' at one electron per two sites. 

The exact solution of this two-site problem for the case $J=0$ has been obtained long ago by Anderson and Hasegawa~\cite{Anderson:1955p675}; here we revisit and generalize this solution to include a nonzero antiferromagnetic interaction between the local moment spins. Our main goal is to study the quantum properties of the KH dimer in the limit of strong Kondo coupling, i.e., $t/J_K, J/J_K\ll 1$, and to demonstrate that the canonical spin wave expansion of Eq.~\eqref{eq:H_KH} correctly reproduces the excitations of the ferromagnet order-by-order in the small parameters of the expansion. To this end, we proceed in two steps. We first determine and analyze the exact solution, and then compute the spin wave excitations of the ferromagnet using the canonical spin wave expansion. We pay particular attention to the instability of the ferromagnet as $J$ is increased and establish to what extent this is well-described by the softening of the spin wave excitation with momentum $\pi$.

\subsection{The Kondo-Heisenberg molecule: exact results \label{ssec:dimer}}

An exact solution of $H$ in Eq.~\eqref{eq:H_KH} can be obtained by computing the matrix elements of \eqref{eq:H_KH} between states that form an invariant subspace. Determining the invariant subspaces amounts to block diagonalizing the full Hamiltonian matrix. The invariant subspaces are labeled by the symmetry quantum numbers of the Hamiltonian, which in the present case are total spin, its and site-exchange parity. 

We define the total spin operator as $\bT =  \bT_1 +  \bT_2$, where $\bT_i = \bS_i + \bs_i$ is the sum of the local moment and electron spin at each site. Hence, $\bT$ is the total spin of the dimer and $\bT_{1,2}$ of each of the two sites. We denote the total spin quantum number as $T$ and the eigenvalue of $T^z$ as $M$. Here $T$ can take $2S$ different values ranging from $1/2$ to $2S+1/2$ and given $T$, $M$ takes $2T+1$ values in the range $-T, \ldots, T$. The two quantum numbers $(T,M)$ label the invariant subspaces.

To determine the dimension of the invariant subspaces consider the state of the electron. The electron can sit at either of the two sites and can form a state of total spin $S\pm 1/2$ with the local moment spin. This implies a four-dimensional invariant subspace for each $(T,M)$, except for the case $T=2S+1/2$, which can only be realized when the electron is in a $S+ 1/2$ state with the local spin. Parity then further partitions the invariant subspace into even and odd sectors. To determine the matrix elements of the Hamiltonian within each subspace, we construct eigenstates of total spin and parity.

The first step in this construction is to form eigenstates of the total spin $\bT_i$ at each site. For instance, for an electron at site $1$ we first form the states $\ket{S\pm\tfrac12,M_1}_1$, which are eigenstates of $\bT_1$ as well as the local Kondo coupling. The explicit expression for these states is given Appendix~\ref{app:dimer}. The second step is to form states with total spin $T$ by coupling a state $\ket{S\pm\tfrac12,M_1}_1$ at site 1 and a state $\ket{S,M_2}_2$ (with no electron) at site 2. This yields eigenstates of total spin given by
\be
\ket{T,M; 1,\pm } = \sum_{M_1,M_2} C^{TM}_{S\pm\frac12,M_1;SM_2}\ket{S\pm\tfrac12,M_1}_1\ket{S,M_2}_2, \label{states-1}
\ee
where $C^{TM}_{S\pm\frac12,M_1;SM_2} $ are the appropriate Clebsch-Gordan coefficients and the sum over is all values of $M_1,M_2$ subject to the constraint $M_1+M_2=M$. Similarly, for an electron sitting at site $2$, we construct states of total spin $T$ as
\be
\ket{T,M;2, \pm } = \sum_{M_1,M_2} C^{TM}_{SM_1;S\pm\frac12,M_2}\ket{S,M_1}_1\ket{S\pm\tfrac12,M_2}_2 \label{states-2}
\ee
The set of states $\{\ket{T,M; 1,\eta } ,\ket{T,M; 2,\eta }  \}$ with $\eta=\pm $ spans the four-dimensional invariant subspace defined by the total spin quantum numbers $(T,M)$. Due to rotational invariance the energies only depend on $T$ and it is therefore sufficient to focus on the highest weight states for which $M=T$. We denote these states simply as $\ket{T; i,\eta }$ ($i=1,2$). Finally, we exploit the invariance under site exchange to form ``bonding'' and ``anti-bonding'' states as
\be
\ket{T; \pm,\eta } = \frac{1}{\sqrt{2}}(\ket{T; 1,\eta }\pm \ket{T; 2,\eta } ), \label{states-symmetrized}
\ee
where $\eta=\pm$ corresponds the two possible total spin states $S\pm1/2$ at each site. It is tempting to conclude that the states $\ket{T; \pm,\eta } $ are even ($+$) and odd ($-$) under site-exchange. This is not correct, however, since it fails to account for the symmetry properties of the Clebsch-Gordan coefficients in \eqref{states-1}. Instead, the eigenvalues under inversion are $\pm e^{i\pi (2S+\eta/2-T)}$, as is explained further in Appendix~\ref{app:dimer}. Note that this implies that the inversion eigenvalues of the states $\ket{T; \pm,\eta } $ alternate as a function of total spin $T$.

 \begin{figure}
	\includegraphics[width=0.9\columnwidth]{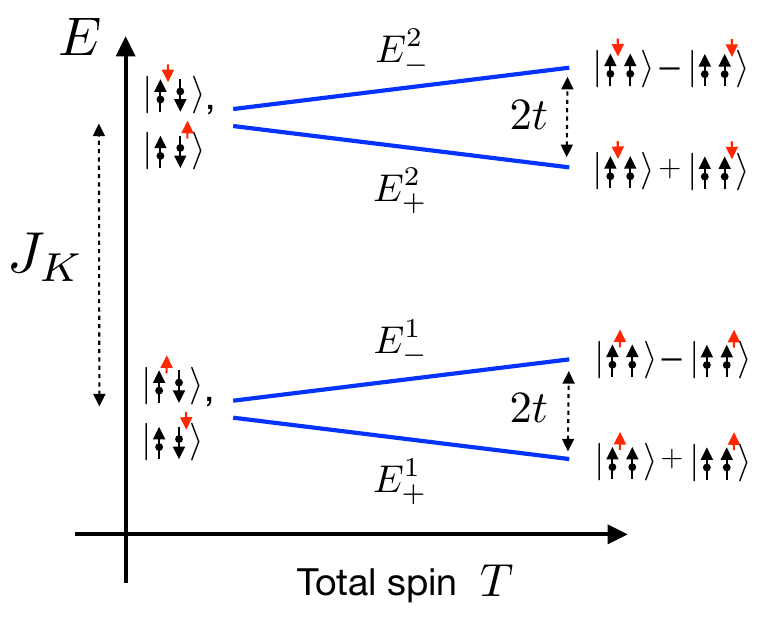}
	\caption{Sketch of the energy spectrum of the KH dimer as given by Eqs.~\eqref{exact-low} and \eqref{exact-high}, shown as a function of total spin $T$. The lower ($E^{1}_{ \pm }$) and upper ($E^{2}_{ \pm }$) two branches are separated by an energy $J_K$ and can be interpreted classically as corresponding to electrons aligned or anti-aligned with the local moment. The eigenstates in the classical limit (i.e., classical spins) are indicated for maximal total spin (i.e., ferromagnet) and minimal total spin (i.e., antiferromagnet). For the case of maximal total spin the bonding and anti-bonding states are separated by an energy $2t$, which is what stabilizes the ferromagnet.}
	\label{fig:dimer_sketch}
\end{figure}

We are now in a position to determine the matrix elements of the Hamiltonian within the invariant subspaces. This is trivial for the special case of maximal total spin $T=2S+1/2$, since the invariant subspaces are non-degenerate. Specifically, the symmetrized bonding and anti-bonding states $\ket{T; \pm, + } $ are automatically eigenstates of the Hamiltonian with eigenvalue $-J_K/2 \mp t +J$. 

For all other values of total spin $T$, we construct the Hamiltonian matrix in the invariant $2\times 2$ subspace defined by the states $\ket{T; \pm,\eta } $. Note that $\pm$ labels the subspaces for given $T$ (and directly relates to parity) and $\eta$ labels the two states in the invariant subspace. We denote the Hamiltonian matrix within each subspace $h_\pm$ and find that it takes the form
\be
h_\pm = \varepsilon + \begin{pmatrix} \chi^z_\pm &  \chi^x_\pm \\ \chi^x_\pm & - \chi^z_\pm  \end{pmatrix}, \label{eq:h_pm}
\ee
where $\varepsilon$, $\chi^z_\pm $, and $\chi^x_\pm$ are given by
\begin{align}
\varepsilon &= \frac{J_K}{4S} +\frac{J (T+1/2)^2}{2S^2} - J(1+1/S), \\
\chi^z_\pm &= -\frac{J_K(1+1/2S)}{2} -\left(J\frac{T+1/2}{2S^2} \pm t \right) \cos \frac{\theta}{2}, \\
\chi^x_\pm &= -\left(J\frac{T+1/2}{2S^2} \pm t \right) \sin \frac{\theta}{2}.
\end{align}
and the parameter $\theta$ is defined by the relation
\be
\cos \frac{\theta}{2} \equiv \frac{T+1/2}{2S+1}.
\ee
This definition is motivated by the fact that in the classical limit ($S \rightarrow \infty$) $\theta$ can be interpreted as the angle between (classical) spins \cite{Anderson:1955p675,Shannon:2001p6371}. Diagonalization of $h_\pm$ leads to two branches $E^{1,2}$ given by
\begin{align}
E^{1}_{ \pm }&=  \varepsilon -\sqrt{(\chi^z_\pm)^2 + (\chi^x_\pm)^2}, \label{exact-low} \\
E^{2}_{ \pm } &=  \varepsilon +\sqrt{(\chi^z_\pm)^2 + (\chi^x_\pm)^2}. \label{exact-high}
\end{align}

We now examine the structure of the energy spectrum. We focus specifically on the regime of strong coupling, when $t/J_K$ is small. A schematic sketch of the energy spectrum in this regime is shown in Fig.~\ref{fig:dimer_sketch}, which serves the purpose of highlighting its essential features. Here we have assumed that $J_K>0$ and $J=0$. The lower two branches, given by $E^{1}_{ \pm }$, and upper two branches, given by $E^{2}_{ \pm }$, are separated by the Kondo energy $J_K$ and can be interpreted as spectral branches corresponding to electrons in a state of on-site spin $S \pm 1/2$, respectively. Classically, these branches can be viewed as describing electrons aligned and anti-aligned with the local moments. (For a discussion of the classical dimer problem see Appendix \ref{app:dimer-classical}.) To get a understanding of the structure of the eigenstates for maximal and minimal total spin, we have indicated the eigenstates of the classical problem when the local moments are fully aligned (ferromagnet) and anti-aligned (antiferromagnet), respectively. 

\begin{figure}
	\includegraphics[width=0.85\columnwidth]{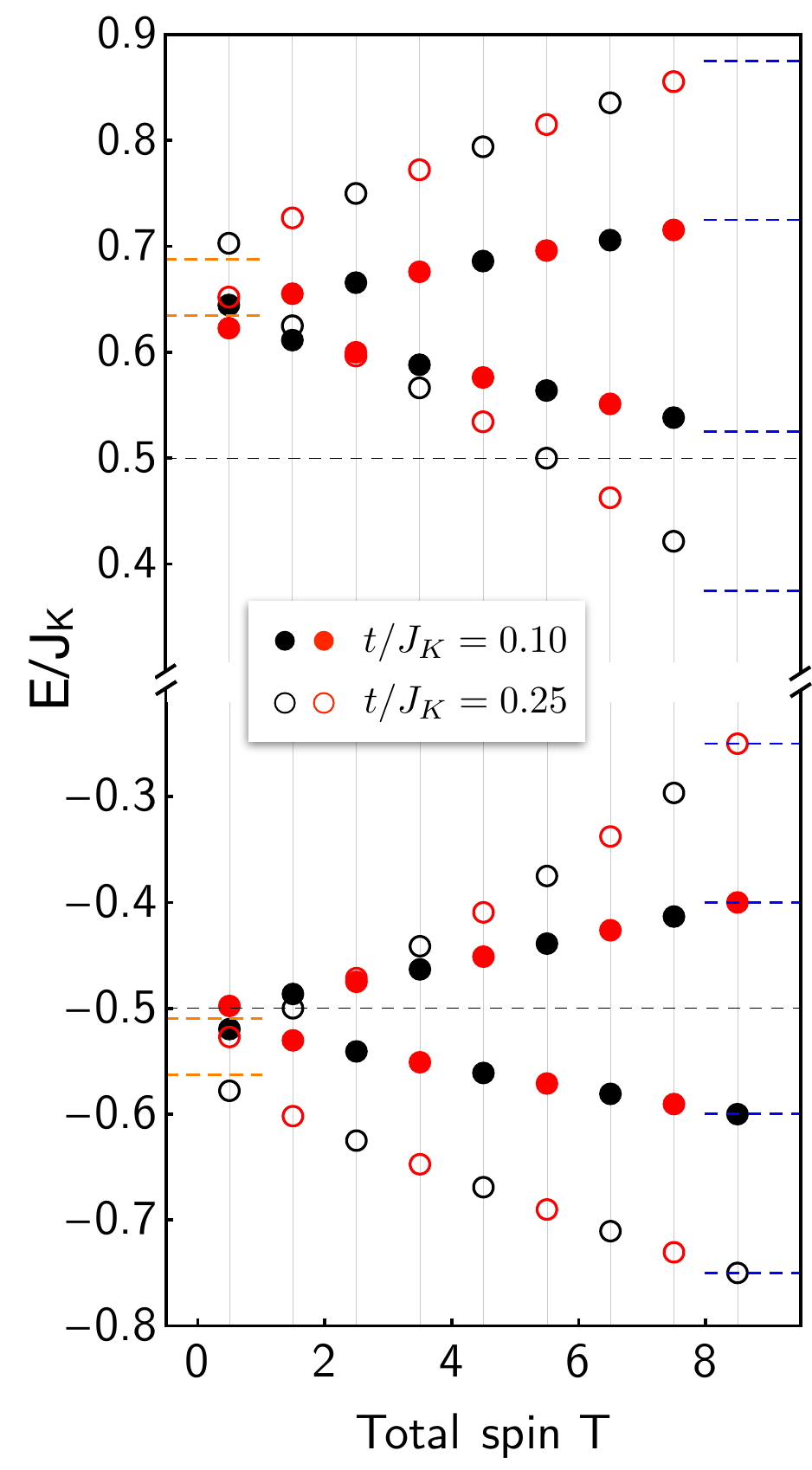}
	\caption{Exact energy levels of the Kondo dimer ($J_K>0$, $J=0$) with $S=4$ spins, as given by Eqs.~\eqref{exact-low} and \eqref{exact-high}. Open (filled) dots correspond to $t/J_K = 0.25$ and $t/J_K = 0.1$, respectively. Black (red) dots indicate site-exchange even (odd) eigenstates. Blue dashed markers on the right vertical axis indicate the value $-J_K/2 \pm t$ (lower two branches given by $E^1_\pm$) and  $J_K(S+1)/2S \pm t$ (upper two branches given by $E^2_\pm$). Orange dashed markers on the left vertical axis indicate the values $-J_K/2 - t^2/J_K$ and $J_K(S+1)/2S + t^2/J_K$.}
	\label{fig:dimer_exact}
\end{figure}

The exact energy levels given by Eqs.~\eqref{exact-low} and \eqref{exact-high} for a dimer of $S=4$ spins is shown in Fig.~\ref{fig:dimer_exact}. Note that $J=0$. The filled and open dots correspond to $t/J_K=0.1$ and $t/J_K=0.25$, respectively, as indicated, and the black and red color coding corresponds to even and odd states under site-exchange. Let us first consider the lower two branches $E^{1}_{ \pm }$ and specifically focus on the cases of maximal and minimal total spin. The exact energy levels of the states with total spin $T=2S+1/2$, here $T=17/2$, are given by $-J_K/2 \pm t$, which coincide with the energies of the classical ferromagnet, as expected. These energies are indicated in Fig.~\ref{fig:dimer_exact} by blue dashed markers. The orange dashed markers, indicated on the left vertical axis, correspond to the energy $  -J_K/2 -  t^2/J_K$, which is the classical energy when the local moments are anti-aligned ($\theta=\pi$). Note that classically the energies are degenerate in case of the antiferromagnet, whereas the energies of the exact quantum problem with $T=1/2$ are split. 

\begin{figure*}
	\includegraphics[width=0.85\textwidth]{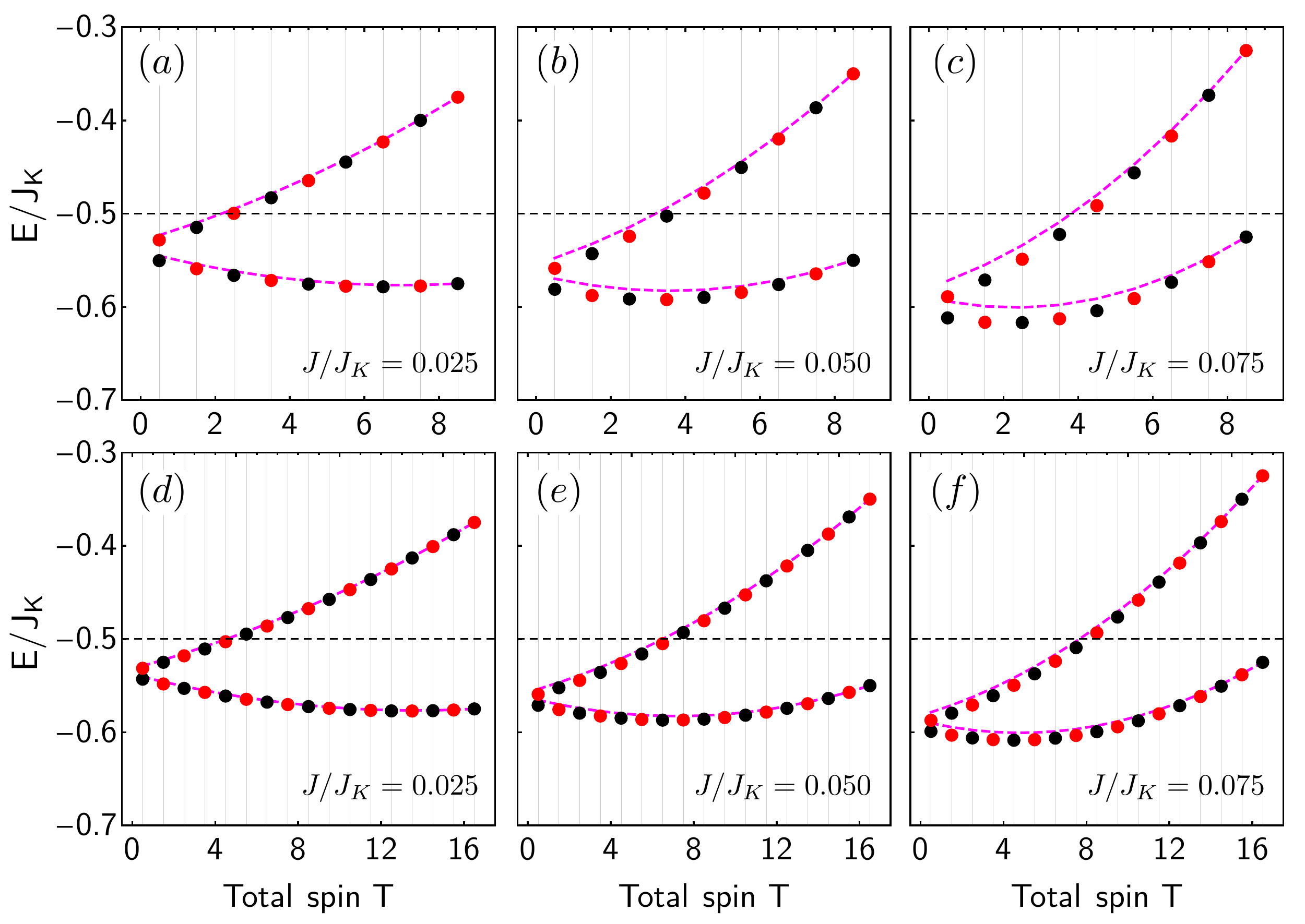}
	\caption{Exact energy levels of the dimer with antiferromagnetic Heisenberg coupling ($J>0$), resolved by total spin $T$. Shown are the lower two branches $E^1_\pm(T)$ of the spectrum [see Eq.~\eqref{exact-low}]. Here $S=4$ in panels (a)-(c) and $S=8$ in panels (d)-(f). For all panels we have chosen $t/J_K=0.1$; values of $J/J_K$ are indicated in the panels.
For comparison with the exact energy levels (shown as bold dots), the magenta dashed lines indicate the corresponding classical approximation, calculated to second in perturbation theory (see Sec.~\ref{ssec:dimer} and Appendix~\ref{app:dimer-classical}). Black and red correspond to parity even and odd solutions, respectively. }
	\label{fig:dimer_afm}
\end{figure*}

Now, consider the upper two branches of the spectrum $E^{2}_{ \pm }$. Here we observe significant deviations from the classical result due to the distinct quantum nature of the $S-1/2$ local spin state. Note that the classical Kondo energy of anti-aligned electron is $J_K/2$ (black dashed line in Fig.~\ref{fig:dimer_exact}), whereas the quantum Kondo energy of a $S-1/2$ local spin state is $J_K(S+1)/2S$. For reference, we indicate the energies $J_K(S+1)/2S \pm  t $ and $J_K(S+1)/2S + t^2/J_K$ by blue and orange dashed markers on the right and left vertical axes, respectively. These are the classical energies of an anti-aligned electron in the ferromagnet and antiferromagnet, shifted by an energy $J_K/2S $. We further note that the upper branches do not have states with total spin $T=2S+1/2$, since such states cannot be formed from two spins with quantum numbers $S$ and $S-1/2$. Hence, the spectrum shown in Fig.~\ref{fig:dimer_exact} demonstrates that the clearest and strongest quantum effects occur in the spectral branches $E^{2}_{ \pm }$, which derive from electrons forming a state of total spin $S-1/2$ with the local moments.

The exact spectrum of the Kondo dimer further confirms the well-known tendency of Kondo lattice models towards ferromagnetism~\cite{Anderson:1955p675}. The dispersion of the spectrum as a function of $T$, which is caused by $t$, the electron kinetic energy scale, gives rise to a ground state with maximal total spin, which is a ferromagnet. This is true both for $J_K>0$, shown in Fig.~\ref{fig:dimer_exact}, and $J_K<0$, in which case $E^{2}_{ \pm }$ form the lower branches and the ground state has total spin $T=2S-1/2$. In the latter case the classical energy is only an approximation of the exact ground state energy. Instead, as noted above, when $J_K>0$ the exact ground state energy of the ferromagnet coincides with the classical ground state energy $-J_K/2 - t$. This is similar to the Heisenberg ferromagnet. In stark contrast with the Heisenberg ferromagnet, however, the spin wave excitations of the ferromagnetic ground state are not exact eigenstates of the Hamiltonian. This will be examined in detail below in Sec.~\ref{ssec:kondo-dimer-FM}, when we compare the lowest energy magnetic excitation in the exact spectrum to lowest energy excitation computed in spin wave theory. 

Next, we examine the spectrum of the dimer in the presence of an antiferromagnetic Heisenberg coupling, focusing in particular on the lower two branches $E^{1}_{ \pm }$. These are shown in Fig.~\ref{fig:dimer_afm} for three different values of $J/J_K$ and for spin $S=4,8$. In all panels we have set $t/J_K=0.1$ and have indicated the classical energies by magenta dashed lines for comparison (see Appendix \ref{app:dimer-classical} for the classical result). The values of $J/J_K$ are chosen such that the ground state is a state of total spin $1/2<T< 2S+1/2$, i.e., a state of neither maximal (ferromagnet) nor minimal (antiferromagnet) total spin $T$. This is due to the competition between the ferromagnetic tendency of the itinerant Kondo model and the antiferromagnetic Heisenberg coupling. We observe that the classical result approximates the quantum problem reasonably for larger spin $S$, especially for $S=8$. We furthermore observe that deviations from the exact quantum result are strongest for smaller total spin $T$, which is exemplified clearest by Fig.~\ref{fig:dimer_afm}(c). This is expected, since the Heisenberg energy of two local moment spins $S$ which together form a state of total spin zero is $-J(S+1)/S$, whereas classically the energy is $-J$. 

For all values of $J/J_K$ shown in Fig.~\ref{fig:dimer_afm} the ferromagnetic state (i.e., the state with $T=2S+1/2$) is no longer the ground state due to the competing Heisenberg coupling $J$. Below, in Sec.~\ref{ssec:kondo-dimer-FM}, we will analyze the instability of the ferromagnet by determining the value of $J$ for which the lowest energy state in the $T=2S-1/2$ sector becomes the ground state, in favor of the ferromagnet.

\subsection{Excitations of the Kondo-Heisenberg ferromagnet \label{ssec:kondo-dimer-FM}}

In this section we consider the KH dimer problem from the perspective of spin wave theory. Specifically, we analyze the magnetic excitations of the two-site KH ferromagnet within the framework of canonical spin wave theory, the spin wave theory for itinerant Kondo magnets developed by the present authors~\cite{our_paper}. The goal is to determine how well canonical spin wave theory describes the lowest energy magnetic excitations of an itinerant ferromagnet. To this end, we compare the energies and wave functions of the spin wave excitations to the exact solution for lowest energy magnetic excitation. Given that canonical spin wave theory is a strong coupling spin wave expansion, our analysis will focus on the strong coupling regime where $J_K$ is the largest energy scale. 

The starting point of spin wave theory is an ordered state defined by a configuration of classical spins. Here, the two-site KH ferromagnet is simply given by two fully aligned classical local moments. As mentioned above in Sec.~\ref{ssec:dimer}, and detailed further in Appendix~\ref{app:dimer-classical}, the ferromagnet is the ground state of the classical KH problem as long as the Heisenberg coupling $J$ is below a critical value (which we will address below). Given the ferromagnetic ground state, the spin wave excitations can be calculated within canonical spin wave theory. The energy and wave function of these spin waves can be compared to the lowest energy magnetic excitation in the $T=2S-1/2$ total spin sector of the quantum problem, assuming that $J_K>0$. Such a comparison is modeled on the approach of Ref.~\onlinecite{Shannon:2001p6371}. 

Canonical spin wave theory is formulated in terms of fermions $f_{i=1,2}$ and bosons $a_{i=1,2}$, which describe, respectively, electrons in a state of total spin $S+1/2$ at site $i$, and fluctuations of that total spin at site $i$. When $J_K>0$, these are the low-energy degrees of freedom of the problem in the strong coupling regime. The general form of the spin wave Hamiltonian describing the dynamics of the low-energy charge ($f_i$) and magnetic ($a_i$) degrees of freedom is presented in Appendix~\ref{app:SWT}, which collects the details necessary for constructing a spin wave theory based on Ref.~\onlinecite{our_paper}. In this section we apply the general expressions summarized in Appendix~\ref{app:SWT} to the dimer problem.

\begin{figure}
	\includegraphics[width=0.85\columnwidth]{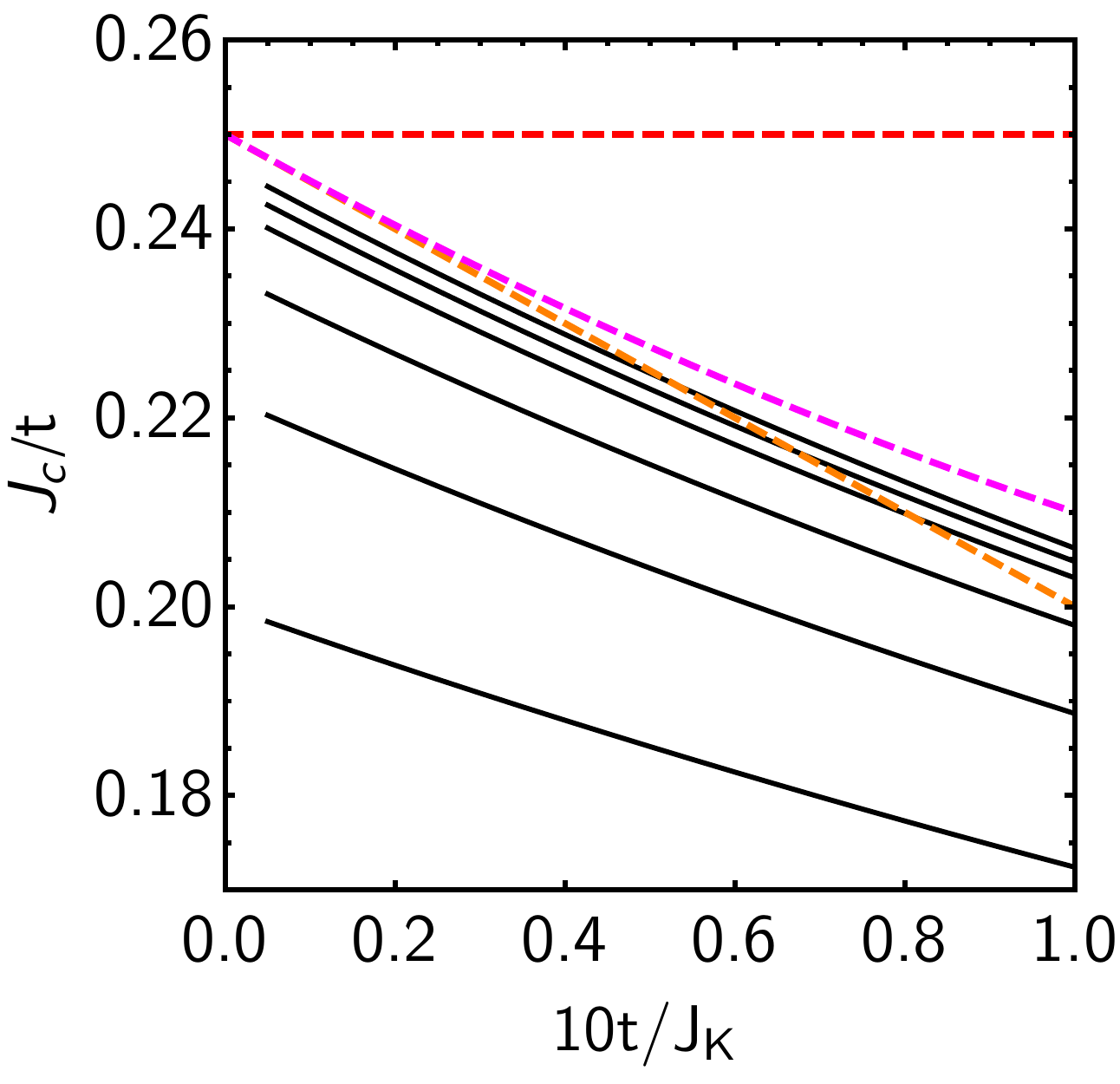}
	\caption{Critical value of the Heisenberg coupling $J_c$ at which the ferromagnet becomes unstable, plotted as a function of $t/J_K$. Black solid lines correspond to the exact value, calculated from Eq.~\eqref{exact-low}, for $S=1,2,4,8,12,20$ (from bottom to top). The colored dashed curves show predictions from linear spin wave theory (i.e., order $1/S$). The red dashed line corresponds to the strong coupling limit; the orange and magenta dashed lines show the result of including $t/J_K$ and $(t/J_K)^2$ corrections to the energy of the spin wave mode, respectively. }
	\label{fig:dimer_instability}
\end{figure}

We proceed by exploiting the site-exchange or parity symmetry of the dimer, which is equivalent to translation, and define the operators
\be
f_{0,\pi} = \frac{1}{\sqrt{2}}(f_{1} \pm f_{2} ), \quad a_{0,\pi} = \frac{1}{\sqrt{2}}(a_{1} \pm a_{2} ). \label{eq:0-pi-operators}
\ee
These correspond to eigenstates of site-exchange. Expressed in terms of these operators, the spin wave Hamiltonian $\mathcal H $ can be written as $\mathcal H = \mathcal H_0+\mathcal H_1$, where $\mathcal H_0$ collects all diagonal terms and is given by
\begin{multline}
\mathcal H_0 =  \varepsilon_0f^\dagger_0 f_0 + \varepsilon_\pi f^\dagger_\pi f_\pi -\frac{2J}{S} a^\dagger_\pi a_\pi  \\ 
+ \left(\frac{t}{2S}- \frac{t^2}{J_KS} -\frac{7t}{32S^2}+ \frac{J}{2S^2}- \frac{2t J}{J_K S^2}  \right) a^\dagger_\pi a_\pi f^\dagger_0f_0  \\
- \left(\frac{t}{2S}+  \frac{t^2}{J_KS} -\frac{7t}{32S^2}- \frac{J}{2S^2}- \frac{2t J}{J_K S^2}  \right) a^\dagger_\pi a_\pi f^\dagger_\pi f_\pi \\
+\frac{t}{32S^2}a^\dagger_0 a_0 (f^\dagger_0 f_0 -f^\dagger_\pi f_\pi )  , \label{eq:H_0}
\end{multline}
with $\varepsilon_{0,\pi} = -J_K/2 \mp t$, and $\mathcal H_1$ collects the relevant off-diagonal terms given by
\be
\mathcal H_1 =  \frac{t}{4S}(a^\dagger_0a_\pi -a^\dagger_\pi a_0 )(f^\dagger_0 f_\pi -f^\dagger_\pi f_0). \label{eq:H_1}
\ee
The Hamiltonian $\mathcal H$ is the result of a spin wave expansion up to order $1/S^2$, including first order corrections in $t/J_K$. In order to obtain the correct energies of the magnetic excitations to order $1/S^2$, we must keep $\mathcal H_1$ and treat it in second order perturbation theory~\cite{Shannon:2001p6371}.

Within canonical spin wave theory, the ground state of the ferromagnet is given by $f^\dagger_0 \ket{\text{FM}}$, where $\ket{\text{FM}}$ is the vacuum for fermions and bosons and represents the classical ferromagnet of local moments. Note that the Heisenberg energy of the two-site local moment ferromagnet is $J$. Therefore, the ground state has energy $E_0 = -J_K/2 - t+J$, which, as mentioned previously, coincides with the energy of the ground state in the quantum problem.

\begin{figure*}
	\includegraphics[width=0.85\textwidth]{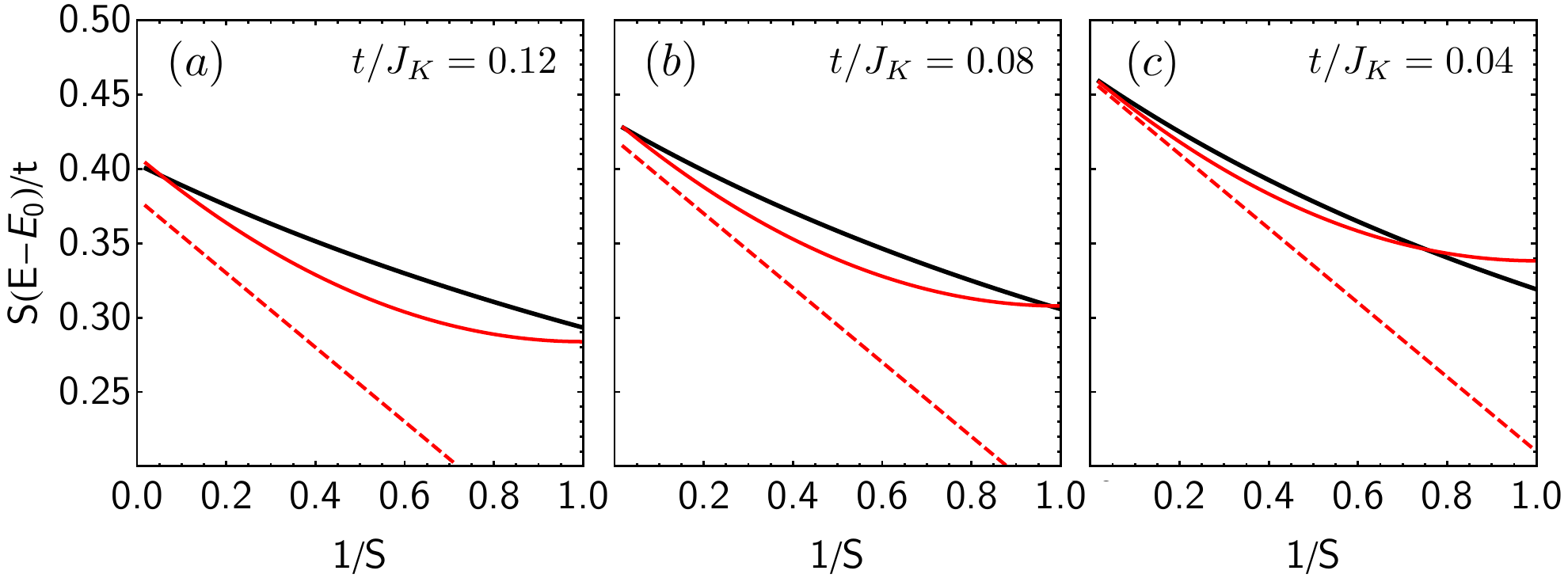}
	\caption{Lowest energy excitation above the ferromagnetic ground state for the Kondo dimer ($J=0$). The difference with the ground state energy $E_0 = -J_K/2-t$ is shown as a function of $1/S$, for three different values of $t/J_K$ (indicated in the panels). }
	\label{fig:dimer_expand}
\end{figure*}  

Our main interest here is in the elementary magnetic excitations given by 
\be
\ket{0,0} =  a^\dagger_0f^\dagger_0 \ket{\text{FM}}, \quad  \ket{\pi,0} = a^\dagger_\pi f^\dagger_0 \ket{\text{FM}},
\ee
for which we determine the energies using the spin wave Hamiltonian $\mathcal H$.
 Consider first the state $\ket{0,0}$. Treating $\mathcal H_1$ in second order perturbation theory cancels the $t/S^2$ contribution from $\mathcal H_0$, and we find that the state $\ket{0,0}$ has energy $E_{0,0} = -J_K/2 - t+J$. Since this is equal to the energy of the ground state, we conclude that $\ket{0,0}$ should be interpreted as a globally rotated ground state, i.e., a state with total spin $T=2S+1/2$ but magnetic quantum number $M=2S-1/2$~\cite{Shannon:2001p6371}. For the state $\ket{\pi,0}$ we find the energy
\begin{multline}
E_{\pi, 0}  =   -\frac12 J_K  -t\left(1 - \frac{1}{2S}+\frac{1}{4S^2} +\frac{t}{J_K S} \right) \\
 + J\left(1 - \frac{2}{S}+\frac{1}{2S^2}-  \frac{2 t}{J_K S^2} \right), \label{eq:E_pi_0}
\end{multline}
which clearly corresponds to a magnetic excitation above the ground state as long as $J$ is small enough for the ground state to remain stable. Below we examine the stability in more detail. 

We first compare the energy $E_{\pi, 0}$ to the exact energy of the first magnetic excitation above the ground state. As discussed in Sec.~\ref{ssec:dimer}, and shown in Fig.~\ref{fig:dimer_exact}, both the ground state and the first magnetic excitation come from the branch $E^1_+= E^1_+(T)$ given by Eq.~\eqref{exact-low}. Specifically, the ground state has energy $E^1_+(2S+1/2) = E_0$ and the first magnetic excitation has energy $E^1_+(2S-1/2)$. Expanding $E^1_+(2S-1/2)$ in $1/S$, $t/J_K$, and $J/ J_K$, and keeping terms to the same order as in Eq.~\eqref{eq:H_0}, we find perfect agreement with $E_{\pi, 0}$. This is our first important result and implies that, as far as the energy is concerned, canonical spin wave theory gives perfect order-by-order agreement with the exact solution. 

We can also compare the state $\ket{\pi,0}$ with the eigenstate wavefunction of the exact first magnetic excitation. The first observation is that $\ket{\pi,0}$ has the correct symmetry: it is odd under site-exchange, as is the exact magnetic excitation with energy $E^1_+(2S-1/2)$ [see Fig.~\ref{fig:dimer_exact} and the discussion following Eq.~\eqref{states-symmetrized}]. The second observation is that $\mathcal H_1$ connects $\ket{\pi,0}$ and $\ket{0,\pi} = a^\dagger_0 f^\dagger_\pi \ket{\text{FM}}$, which implies that the correct eigenstate of $\mathcal H$ with energy $E_{\pi, 0} $ up to first order in $1/S$ is $\ket{\pi,0} + \ket{0,\pi}/8S$. (Recall that $E_{\pi, 0} $ was determined by treating $\mathcal H_1$ in second order perturbation theory.) To proceed, we then express the eigenstate in terms of the original degrees of freedom, i.e., operators corresponding to the constituent electron and local moment spins. Appendix \ref{app:SWT} provides further details on the definition of $f_i$ and $a_i$ in terms of the original degrees of freedom (i.e., the electron operators $c_{i\up}$ and $c_{i\down}$, and the bosonic spin flips of the local moment spin). Using these expression we find
\begin{align}
& \left(a^\dagger_\pi f^\dagger_0+\frac{1}{8S}a^\dagger_0 f^\dagger_\pi  \right) \ket{\text{FM}} \\
& = \left(1-\frac{1}{8S}+\frac{t}{2J_KS} \right)a^\dagger_\pi c^\dagger_{0\up } \ket{\text{FM}} \nonumber \\
&\quad - \left(\frac{1}{4S}-\frac{t}{2J_KS} \right)a^\dagger_0 c^\dagger_{\pi\up } \ket{\text{FM}}+ \frac{1}{2\sqrt{S}} c^\dagger_{\pi\down} \ket{\text{FM}}, \label{eq:E_pi_0-state}
\end{align}
where on the right hand side the bosons represent fluctuations of the local moment spin only. The right hand side can now be compared with the exact eigenstate obtained from the solution of \eqref{eq:h_pm} with energy $E^1_+(2S-1/2)$. Using equalities such as $\ket{S,S-1}_1\ket{S,S}_2 = a^\dagger_1\ket{\text{FM}}$, and expanding the exact eigenstate in $1/S$ and $t/J_K$, we find full agreement with \eqref{eq:E_pi_0-state}. This shows that canonical spin wave theory correctly captures both the energies and eigenstates of the full quantum KH dimer problem order-by-order in the spin wave expansion parameters. 

Next, we turn to an analysis of the instability of the ferromagnet. In general, within spin wave theory, an instability is signaled by the appearance of a zero mode in the spin wave spectrum. Here we therefore examine the energy of the excitation $\ket{\pi,0}$ and determine when it vanishes. We focus on the linear spin wave terms of $E_{\pi, 0}$ in \eqref{eq:E_pi_0}, which are terms of order $1/S$. In the strict strong coupling limit, the linear spin wave energy equals $t/2 -2J$, which yields a critical value of $J_c = t/4$. When including the first order $t/J_K$ correction, the critical value of $J$ decreases and becomes $J_c = t/4 - t^2/ 2J_K$. These values are shown as a function of $t/J_K$ in Fig.~\ref{fig:dimer_instability}, where the horizontal dashed red line corresponds to the strong coupling limit and the dashed orange curve includes the $t/J_K$ corrections. The magenta curve is obtained from including the next-order $t^2/J^2_K$ correction to the linear spin wave energy, and shows that this correction increases the energy. For comparison, the black curves show the critical value $J_c$ obtained from the exact solution, which is found by determining when $\Delta E =  E^1_+(2S-1/2) - E_0$ changes sign from positive to negative. Different curves correspond to different values of $S$ and show that spin wave theory becomes more accurate as $S$ increases, as expected. 

The analysis of the instability of the ferromagnet shows that linear spin wave theory of Kondo lattice magnets is not exact, but approaches the exact result as $S$ increases. In Fig.~\ref{fig:dimer_expand} we further examine how well our spin wave expansion approximates the magnetic excitations of a Kondo lattice ferromagnet. Shown in Fig.~\ref{fig:dimer_expand} is a comparison between the exact energy of the lowest magnetic excitation and the energy obtained within spin wave theory. The energies are shown as a function of $1/S$ and for different values of $t/J_K$. The dashed red curve corresponds to $E_{\pi, 0}  $ as given by Eq.~\eqref{eq:E_pi_0} with $J=0$ and the solid red curve corresponds to 
\be
S(E_{\pi, 0} - E_0)/t  =  \frac{1}{2} -\frac{t}{J_K } +\frac{2t^2}{J^2_K } - \frac{1}{4S}+\frac{1}{8S^2},
\ee
which includes higher order corrections in $1/S$ and $t/J_K$. The black line shows the exact energy.



\section{Kondo-Heisenberg chain in one dimension: Exact results \label{sec:KHChain-exact}}

In this section we turn to the analysis of the KH chain in 1D. The Hamiltonian of the KH model in 1D is an immediate generalization of Eq.~\eqref{eq:H_KH} and is given by
\be
H =  \sum_{ij} t_{ij} c^\dagger_{i} c_{j}  -\frac{J_K}{S} \sum_{i}  \bS_i\cdot \bs_i+\frac{1}{2S^2} \sum_{ij}J_{ij} \bS_i\cdot \bS_j. \label{eq:1D-KH}
\ee
We choose the hopping $t_{ij}$ and the Heisenberg exchange coupling $J_{ij}$ to be between nearest neighbors only. More specifically, we take $t_{ij}=-t$ and $J_{ij}=J$ when $\langle ij\rangle$ are nearest neighbors and zero otherwise. We furthermore consider a lattice of $N$ spins and a single electron. 

The 1D Kondo chain with a single electron has previously been studied to gain insight into the phase diagram of the Kondo model in the limit of (vanishingly) small electron density~\cite{Ueda:1991p167,Sigrist:1991p2211,Sigrist:1992p175,Tsunetsugu:1997p809}. The special case of a single electron provides an appealing toy model for the extremely dilute limit, since it can be proven that the ground state is a ferromagnet and an equation for the energy of the ground state can be obtained analytically~\cite{Sigrist:1991p2211}. 

Here we revisit the problem of the 1D Kondo model in a more general setting, where we include an antiferromagnetic Heisenberg exchange coupling between the local moments and---most importantly for our purposes---consider a general local moment spin $S$~\cite{Henning:2012p085101,Masui:2022p014411}. As in the case of the dimer, our goal is to examine the lowest energy excitations of the KH ferromagnet, and in particular to compare the excitation energies obtained from spin wave theory to the exact solution of the Schr\"odinger equation. This not only provides further insight into the nature and adequacy of the spin wave expansion, but also sheds further light on the fundamental distinction between an itinerant Kondo ferromagnet and a Heisenberg ferromagnet. In the case of the latter, the spin wave excitations are exact eigenstates of the Hamiltonian. This is not the case for itinerant Kondo ferromagnets and the analysis presented in the remaining sections will demonstrate how well spin wave theory captures the exact result. 


Our analysis of the KH chain proceeds in three steps: in this section, we determine the full exact solution in specific sectors of total spin, the sectors with total spin $T=NS\pm 1/2$. In Sec.~\ref{sec:SWexp} we then apply canonical spin wave theory to the KH ferromagnet, assuming a ferromagnetic Kondo coupling ($J_K>0$ in our convention). Finally, in Sec.~\ref{sec:SWexp} we focus on the case of antiferromagnetic Kondo coupling. In this case the electron forms a bound state with spin flip excitations in the ground state and we will show that the spin wave expansion naturally captures this spin polaron state.

\subsection{Ferromagnetic ground state in the $J=0$ limit \label{ssec:1D_GS}}

We first establish that in the $J=0$ limit the ground state of the single-electron Kondo lattice model of Eq.~\eqref{eq:1D-KH} is a ferromagnet. The argument follows in a straightforward manner from the previously given proof for the case $S=1/2$~\cite{Sigrist:1991p2211,Tsunetsugu:1997p809}. 

It is important to distinguish ferromagnetic ($J_K>0$) and antiferromagnetic ($J_K<0$) Kondo coupling. Consider first the case $J_K>0$. Ignoring the spin-flip piece of the Kondo coupling and keeping only the Ising term, it is easy to see that the ground state is given by $\ket{\Phi} =N^{-1/2}\sum_{i} c^\dagger_{i\up}  \ket{\text{FM}}$ with energy $ - J_K/2-2t $, where $\ket{\text{FM}} = \otimes_j \ket{S,S}_j$ is the fully polarized state of local spins. Since it has total spin $T=NS+1/2$, this state remains the exact ground state when the spin-flip terms are reinstated~\cite{Sigrist:1991p2211}. Note that the ground state is unique up to a $2T+1 $-fold degeneracy. 
 
The situation is different when $J_K<0$, since now the lowest energy state in the Ising limit, given by $N^{-1/2}\sum_{i} c^\dagger_{i\down}  \ket{\text{FM}}$, is not an eigenstate of the Hamiltonian with isotropic Kondo coupling. To show that the ground state nonetheless has total spin $T= NS-1/2$, we straightforwardly adapt the proof of Ref.~\onlinecite{Sigrist:1991p2211}, which is based on the Perron-Frobenius theorem. It can be shown that, when constructed in a properly chosen basis, the Hamiltonian matrix within an invariant subspace defined by the eigenvalue of $T^z$ (i.e., the $z$-component of total spin) has two properties: ({\it i}) it only has nonpositive off-diagonal matrix elements and ({\it ii}) successive application of the Hamiltonian eventually connects all states within the invariant subspace. Given these properties, the Perron-Frobenius theorem then states that the lowest energy eigenstate within each subspace has a strictly positive wave function and is unique. These two implications can be used to demonstrate that the ground state is also an eigenstate of $\bT^2$ and has total spin quantum number $T=NS-1/2$. 

These arguments establish that the ground state is a ferromagnet for either sign of $J_K$ when $J=0$. The effect of $J$ can then be determined by examining the stability of the ferromagnet as $J$ is increased. For the case $J_K>0$, this is achieved by solving the 1D KH model for finite $J$ in the sector of total spin $T=NS-1/2$, since this total spin sector contains the lowest energy magnetic excitations above the ground state, i.e., the spin waves. When the difference between the energy of (one of) these excitations and the ground state vanishes for some value of $J$, the ferromagnetic becomes unstable. For the case $J_K<0$, determining the instability of the ferromagnet would require solving for the magnetic excitations in the $T=NS-3/2$ total spin sector. This is a more complicated problem which we do not attempt to address here. 

Before solving for the energies and eigenstates in the $T=NS-1/2$ sector, let us for completeness state the full solution in the sector of maximal total spin $T=NS+1/2$~\cite{Shastry:1981p5340,Sigrist:1991p2211,Henning:2012p085101}. Here the eigenstates are nondegenerate and are labeled by momentum $p$: $\ket{\Phi_p} =c^\dagger_{p \up}  \ket{\text{FM}}$. They have energy $E_p = \varepsilon_p - J_K/2 + E^J_c$, where
\be
\varepsilon_p = -2t\cos p, \quad E^J_c =  \frac12 \sum_{ij} J_{ij} = \frac12 z J N, \label{eps_p}
\ee
are the electron kinetic energy and the classical energy of the Heisenberg ferromagnet, respectively. For a general group theory analysis of the spectrum of the 1D chain irrespective of $T$, see Appendix~\ref{app:1D-group-theory}.

\subsection{Exact solution for total spin $T=NS-1/2$ \label{ssec:1D_exact}}

We follow Ref.~\onlinecite{Sigrist:1991p2211} (see also Ref.~\onlinecite{Henning:2012p085101}) and consider a general wave function in the subspace of states with $T^z$ eigenvalue $ NS-1/2$ given by
\be
\ket{\Psi} = \sum_{i}\bigg(\Psi^i c^\dagger_{i\down} + \frac{1}{\sqrt{2S}}\sum_j\Psi^i_jc^\dagger_{i\up}S^-_j \bigg) \ket{\text{FM}}. \label{eq:Psi-def}
\ee
This subspace includes unwanted states with total spin $T= NS+1/2$. To project these states out, we impose the constraint $T^+ \ket{\Psi} =0$, where $T^+ =  \sum_i (s^+_i + S^+_i)$ is the raising operator of total spin, leading to
\be
 \Psi^i + \sqrt{2S}\sum_j \Psi^i_j =0. \label{constraint}
\ee
The Schr\"odinger equation $( H - E^J_c ) \ket{\Psi}  = E\ket{\Psi}$ must be solved subject to this constraint. Note that here we have subtracted the classical energy of the Heisenberg ferromagnet, defined in \eqref{eps_p}, which is just an overall constant. 

The Schr\"odinger equation gives rise to two coupled equations for the wave function components. These are given by
\begin{align}
E \Psi^i & =   \sum_j t_{ij}\Psi^j + \frac12 J_K \Psi^i - \frac{J_K}{\sqrt{2S}}\Psi^i_i , \label{eq:Psi_i}\\
 E \Psi^i_j & = \sum_k t_{ik}\Psi^k_j-  \frac{J_K}{\sqrt{2S}} \delta_{ij}\Psi^i-\frac{J_K}{2S} (S - \delta_{ij})\Psi^i_j  \nonumber \\
 &\qquad  + \frac{1}{S}\sum_k J_{kj}( \Psi^i_k- \Psi^i_j ) \label{eq:Psi_ij}.
\end{align}
To solve these equations, we exploit translational invariance and take the Fourier transform. We define the Fourier transform of the wave function as
\begin{align}
\widetilde \Psi^p & = \frac{1}{\sqrt{N}} \sum_i \Psi^i e^{-i p r_i}, \\
\widetilde \Psi^p_q &= \frac{1}{N} \sum_{ij} \Psi^i_j e^{-i p r_i}e^{-i q (r_i-r_j)}.
\end{align}
where $p$ is the conserved (center-of-mass) momentum. The Fourier transformed Schr\"odinger equations take the form
\begin{align}
0& =  \left(\varepsilon_p + \frac12J_K -E\right)\widetilde \Psi^p -  \frac{J_K}{\sqrt{2S}} \widetilde\Sigma^p ,\label{eq:Psi_p} \\
 0  & = \left(\varepsilon_{p+q} - \frac12J_K - \frac{1}{S} J  \gamma_q -E\right)\widetilde \Psi^p_q -\frac{J_K }{\sqrt{2SN}}\widetilde \Psi^p  \nonumber \\
 &\hskip1.8in\relax +\frac{J_K}{2S\sqrt{N}}  \widetilde\Sigma^p , \label{eq:Psi_pq}
\end{align}
where we have defined 
\be
\widetilde\Sigma^p \equiv \frac{1}{\sqrt{N}}\sum_q \widetilde \Psi^p_q, \quad \gamma_q \equiv 2(1-\cos q).
\ee
We note in passing that $J \gamma_q $ corresponds to the familiar dispersion of a Heisenberg ferromagnet in 1D. 

The two equations \eqref{eq:Psi_p} and \eqref{eq:Psi_pq} are diagonal in the center-of-mass momentum $p$, but couple different values of $q$. These equations can thus be solved for a fixed value of $p$ under the constraint $\widetilde\Psi^p + \sqrt{2SN}\widetilde\Psi^p_0 = 0$,
which is the Fourier transform of Eq.~\eqref{constraint}. From \eqref{eq:Psi_p} and \eqref{eq:Psi_pq} we obtain a relationship between the wave function components given by
\be
\widetilde{\Psi}_q^p = -\frac{1}{\sqrt{2SN}}\frac{\varepsilon_p - \frac{1}{2}J_K-E}{\varepsilon_{p+q}-\frac{1}{2}J_K-\frac{1}{S}J\gamma_q-E}\widetilde{\Psi}^p \label{eq:PsiPsi}
\ee
which can be used to derive implicit equation for the energies $E$ (see Appendix \ref{app:1DJ} for details). We find
\be
\frac{\varepsilon_p+\frac{1}{2}J_K-E}{\varepsilon_p-\frac{1}{2}J_K-E}=\frac{1}{SN}\sum_q \frac{\frac12 J_K}{E -\varepsilon_{p+q}+\frac{1}{2}J_K+\frac{1}{S}J\gamma_q} \label{eq:ImpEqJ}
\ee
There also exist solutions of \eqref{eq:Psi_p} and \eqref{eq:Psi_pq} not described by \eqref{eq:PsiPsi} and \eqref{eq:ImpEqJ}. These are solutions for which $\widetilde{\Psi}^p =0 $. The construction of these solutions is described in Appendix \ref{app:1DJ}. The special case $J=0$, when the Heisenberg coupling is absent, is considered in detail in Appendix \ref{app:1DJ0}.

\begin{figure}
     \centering
     \includegraphics[width=\columnwidth]{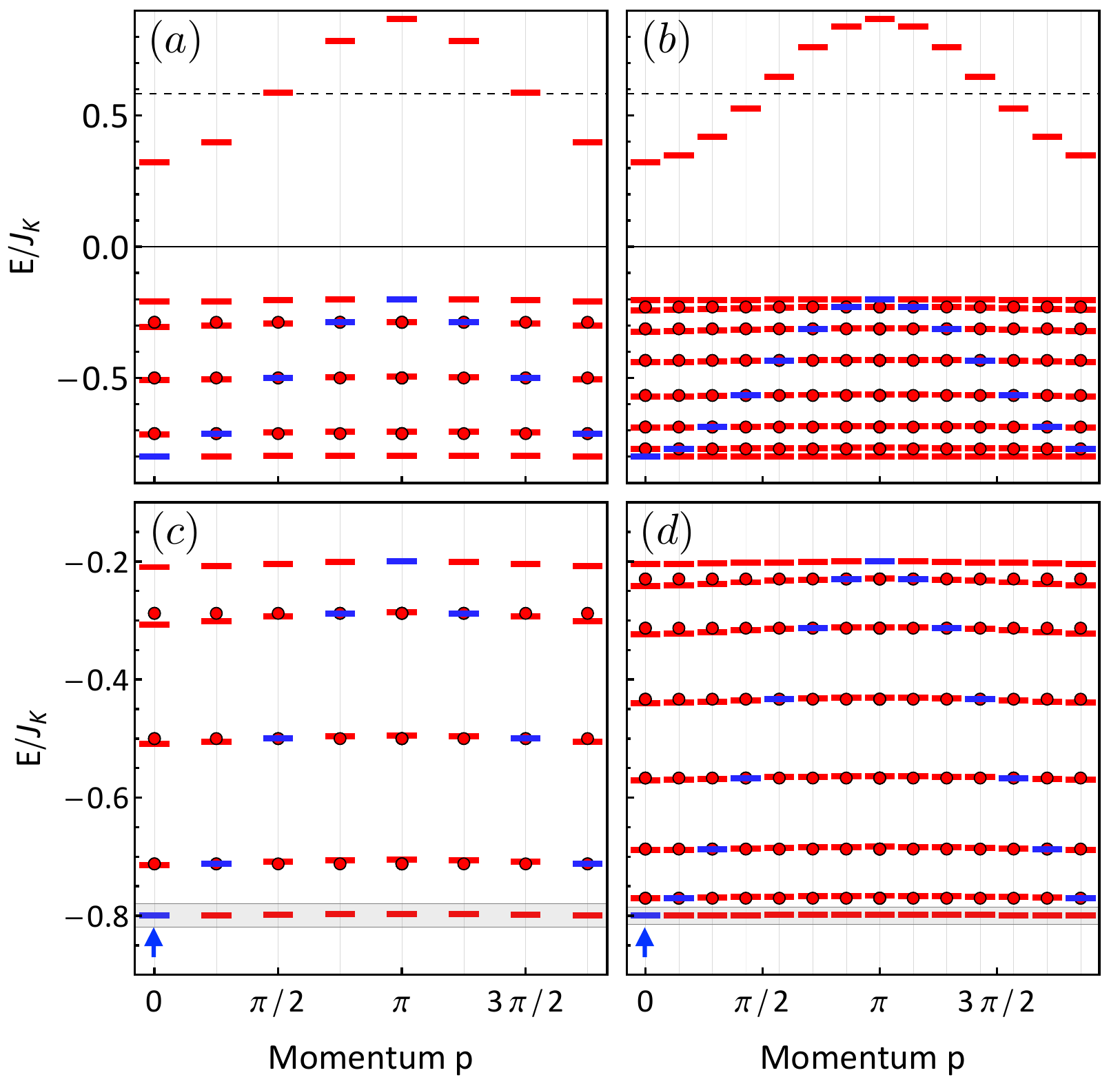}
     \caption{Exact energy levels of the 1D Kondo chain [$N=8$ in panels (a) and (c); $N=14$ in panels (b) and (d)] shown as function of center-of-mass momentum $p$. In all panels we have chosen $S = 6 $, $t/J_{K}=0.15$, and $J=0$. Panels (a) and (b) show the full spectrum, and panels (c) and (d) zoom in on the two-particle-continuum. Red bars correspond to solutions of Eq.~\eqref{eq:ImpEqJ}; red dots correspond to special solutions not given by \eqref{eq:ImpEqJ} (see Appendices \ref{app:1DJ} and \ref{app:1DJ0}). For reference, the blue bars show the energy levels of the $T = NS + 1/2$ total spin sector given by $E_p = \varepsilon_p - J_K/2 $, see Eq.~\eqref{eps_p}. In panels (c) and (d) the ground state is indicated by a blue arrow and the band of lowest energy magnetic excitations is highlighted in grey. In panels (a) and (b) the horizontal black dashed line corresponds to $J_K(S+1)/2S$.}
    \label{fig:KH_1D_exact}
\end{figure}

We proceed by examining the obtained energy spectrum of the 1D chain (in the sector of total spin $T=NS-1/2$). In Fig.~\ref{fig:KH_1D_exact} we show the exact spectrum for chains of length $N=8 $ and $N=14$, choosing a ferromagnetic Kondo coupling $J_K>0$ and setting the Heisenberg coupling $J$ to zero. We have further chosen $S=6$ and $t/J_K = 0.15$, since our interest is in the strong coupling regime. The red bars and dots correspond energy levels obtained from Eq.~\eqref{eq:ImpEqJ} and the special solutions not captured by Eq.~\eqref{eq:ImpEqJ}, respectively. For comparison, the blue bars indicate the energy levels in the total spin $T=NS+1/2$ sector. Note that the ground state, which lies in this sector and is indicated by blue arrows in panels (c) and (d), has momentum $p=0$ and energy $E_0 = -J_K/2 -2t$. 

Panels (a) and (b) of Fig.~\ref{fig:KH_1D_exact} show full spectrum. It exhibits two main features. The first feature is a manifold of energy levels at negative energies, which corresponds to the two-particle continuum of charge and magnetic excitations above the ground state. The latter is emphasized by the superimposed solutions of the $T=NS+1/2$ sector marked in blue. The states in this two-particle continuum can be thought of as electrons forming a state of total spin $S+1/2$ with the local moment. Panels (c) and (d) zoom in on the two-particle continuum, showing that the energy levels marked by red bars have weak dispersion with width $\sim t/SN$. Note further that the width of the two-particle continuum itself is $4t$, equal to the bandwidth of the electrons, and that the center is at energy $-J_K/2$, the on-site Kondo energy of an electron in a state of total spin $S+1/2$. In panels (c) and (d) we have indicated the band of lowest energy magnetic excitations in grey, which will be examined in detail in the next section. 

The second feature is a single isolated band at positive energy. The states forming this band are spin polaron states and can be understood as bound states of a spin-down electron and spin-flip excitations, i.e., a spin-down electron dressed with a cloud of spin-flip excitations~\cite{Sigrist:1991p2211,Tsunetsugu:1997p809}. This interpretation is clearly suggested by the structure of the wave function in Eq.~\eqref{eq:Psi-def}. The spin polaron band has bandwidth $~4t$ and sits at energy $J_K(S+1)/2S$, which is indicated by a dashed horizontal line. This emphasizes an alternative way to understand the spin polaron band, namely as describing the dispersion of an electron forming a state of spin $S-1/2$ with the local moments. The latter interpretation will be considered in more detail in Sec.~\ref{sec:1D-polaron}, where we analyze the structure of the spin polaron states by computing the spin-spin correlation function both exactly and within the framework of the spin wave expansion.

\section{Kondo-Heisenberg chain in one dimension: Spin wave expansion \label{sec:SWexp}}

In this section we analyze the 1D KH chain from the perspective of spin wave theory. As in the case of the dimer, we apply the strong coupling spin wave expansion of Ref.~\onlinecite{our_paper} and specifically focus on two aspects of the spin wave theory.  First, we analyze the spin wave excitations of the ferromagnetic ground state (assuming $J_K>0$) and compare these to the exact solution in the $T=NS-1/2$ total spin sector. We then address the instability of the ferromagnet as antiferromagetic Heisenberg coupling $J$ is increased. 


\subsection{Spin wave excitations of the ferromagnet \label{ssec:spin-waves}}

\begin{figure}
    \centering
    \includegraphics[width=0.75\columnwidth]{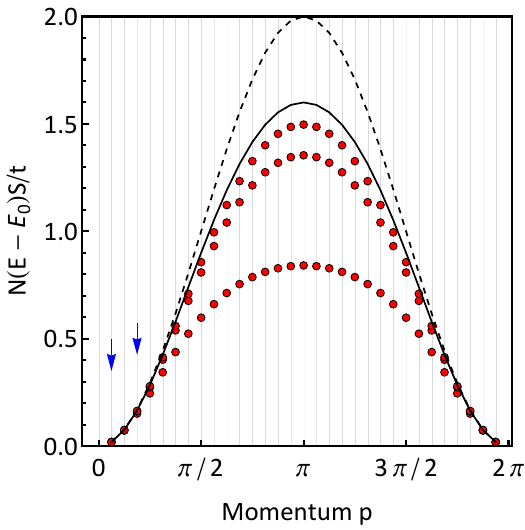}
    \caption{Band of lowest energy magnetic excitations of the 1D Kondo ferromagnet ($J = 0$) for a chain of length $N=32$. Red dots correspond to the exact energies, obtained from \eqref{eq:ImpEqJ} for three different values of $S$ ($S = 5, 20, 40$ from bottom to top) and $t/J_K = 0.05$, shown as a function of center-of-mass momentum $p$. The black curves show the energies obtained from linear spin wave theory, as described by the $1/S$ terms in \eqref{eq:E_q-SW}. The dashed black curve corresponds to the strong coupling limit given by $t(1+c_p)/NS$. The solid black curve corresponds to Eq.~\eqref{eq:E_p-LSWT}, which includes the $t/J_K$ correction to strong coupling limit. }
    \label{fig:LSWT-comp}
\end{figure}

\begin{figure}
    \centering
    \includegraphics[width=0.75\columnwidth]{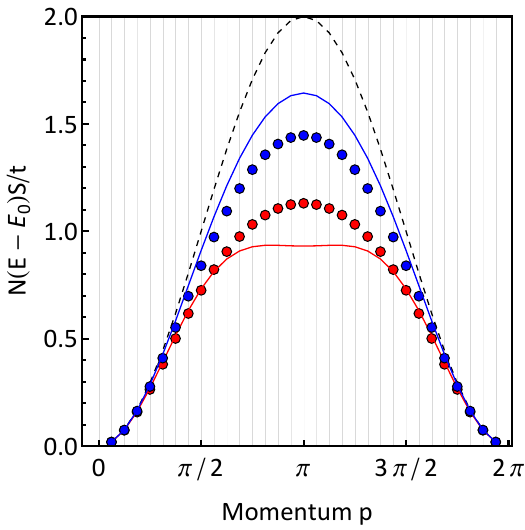}
    \caption{Lowest energy magnetic excitations of the $N=32$ 1D Kondo ferromagnet ($J=0$) with $t/J_K = 0.05$ (see also Fig.~\ref{fig:LSWT-comp}). Red and blue dots correspond to the exact energies obtained from \eqref{eq:ImpEqJ} for $S=10$ and $S=30$, respectively. The dashed black curve corresponds to the linear spin wave energies in the strong coupling limit, as in Fig.~\ref{fig:LSWT-comp}. The red ($S=10$) and blue ($S=30$) curves show the spin wave energies including $1/S$ corrections, as given by \eqref{eq:E_q-SW}. }
    \label{fig:SWT-comp}
\end{figure}

Within canonical spin wave theory, the ferromagnetic ground state is given by $f^\dagger_0 \ket{\text{FM}}$, where $f^\dagger_0   =N^{-1/2}\sum_i f^\dagger_i $ creates an zero momentum superposition of electrons in a $S+1/2$ state, and $\ket{\text{FM}}$ is the classical local moment ferromagnet (and the vacuum for fermions and bosons). The energy of the ground state is $E_0= -J_K/2-2t$.
Our interest here is in the spin wave excitations above the ferromagnetic ground state. These spin wave excitations are part of the two-particle continuum of charge and magnetic excitations, which, within spin wave theory, are given by the states 
\be
\ket{p, q} \equiv a^\dagger_q  f^\dagger_{p-q} \ket{\text{FM}}. 
\ee
Note that, as is usual in a spin wave theory, the $q=0$ states are simply uniform rotations of the states $ f^\dagger_{p} \ket{\text{FM}}$ and thus have the same energy. In particular, the $\ket{0,0}$ state is a uniform rotation of the ferromagnetic ground state. 

The spin wave excitations of interest are given by $\ket{p,p} = a^\dagger_p  f^\dagger_{0} \ket{\text{FM}}$. The energy of these states can be calculated using the canonical spin wave Hamiltonian; details are provided in Appendix~\ref{app:SWT}. We find the energies as
\begin{multline}
E_p = E_0+\frac{t}{SN}\left[1-c_p -\frac{2t}{J_K}\left(1-c_p\right)^2 \right. \\
    \left.+\frac{1}{8S}\left(4c_p-3\right) - \frac{1}{8SN}\sum_{q\neq 0} \frac{\left(2c_p-c_{q}-1\right)^2}{1-c_{q}}\right] \\
    -\frac{J}{S}\left(1-c_p\right)\left[2 - \frac{1}{SN}+ \frac{4t}{J_KSN}\left(1-c_p\right)\right] ,\label{eq:E_q-SW}
\end{multline} 
where we have defined $c_p \equiv \cos p$. These energies obtained from the spin wave expansion can be compared to the energies of the exact magnetic excitations of the ferromagnet. Specifically, the energies of Eq.~\eqref{eq:E_q-SW} should be compared to the band of pure magnetic excitations in the two-particle continuum, which are highlighted in panels (c) and (b) of Fig.~\ref{fig:KH_1D_exact} by a light grey box.

We first consider this comparison for the case $J=0$. For small chains with $N=3,4,6$, exact closed form expressions for the energies of the magnetic excitations can be obtained from Eq.~\eqref{eq:ImpEqJ} (with the help of \textsc{Mathematica}). Expanding the exact expression in $1/S$ and $t/J_K$ yields exact order-by-order agreement with Eq.~\eqref{eq:E_q-SW}, which confirms and further cements the conclusion that the canonical spin wave expansion correctly captures the features of the exact solution order-by-order in the expansion parameters.

To further examine the spin wave energies, consider Fig.~\ref{fig:LSWT-comp}, where we show the exact energies as a function of momentum $p$ for a chain of $N=32$ sites and three values of $S$. The exact energies are represented by red dots. For comparison, we also show the spin wave energies in linear spin wave theory, both in the strong coupling limit (dashed line) and including first order $t/J_K$ corrections (solid line). In the strong coupling limit, the spin wave dispersion \eqref{eq:E_q-SW} is simply given by $E_p - E_0 \simeq t(1-c_p)/NS$, which is equivalent to the linear spin wave dispersion of a Heisenberg ferromagnet with effective exchange coupling $t/2N$. This is in agreement with previous studies of itinerant Kondo lattice ferromagnets, which found that the effective Heisenberg coupling is equal to the average electron kinetic energy per bond. Since in this case we consider a system with a single electron, the effective coupling scales as $\sim 1/N$.

When the $t/J_K$ corrections are included, the linear spin wave dispersion becomes 
\be
E_p - E_0\simeq \frac{t}{SN}\left[1-c_p -\frac{2t}{J_K}\left(1-c_p\right)^2 \right], \label{eq:E_p-LSWT}
\ee
which may be rewritten as $NS(E_p - E_0)/t \simeq (1-4t/J_K)(1-c_p)+t(1-c_{2p})/J_K$. The latter expression has the form $2 J_1 (1-c_p)+2  J_2(1-c_{2p})$ and is thus formally equivalent to the spin wave dispersion of a Heisenberg ferromagnet with nearest and next-nearest neighbor ferromagnetic exchange couplings $J_1= t(1-4t/J_K)/2N$ and $J_2 = t^2/2NJ_K$. We therefore find that the $t/J_K$ corrections reduce the effective nearest neighbor exchange coupling and furthermore introduce an additional effective next-nearest neighbor coupling. Note that the small momentum $p\rightarrow 0$ spin wave excitations are not affected by these corrections. Indeed, expanding in small momentum $p$ we find $E_p - E_0 \approx t p^2 /2NS$, which is reflected in Fig.~\ref{fig:LSWT-comp} by the behavior of the dashed and solid black lines at small momentum. 

\begin{figure}
     \centering
     \includegraphics[width=0.75\columnwidth]{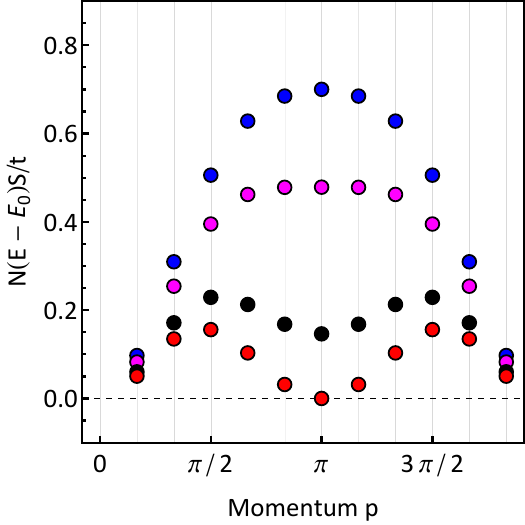}
     \caption{Band of exact lowest energy magnetic excitations of the 1D Kondo-Heisenberg ferromagnetic (i.e., antiferromagnetic $J$ included). Energy levels calculated using Eq.~\eqref{eq:ImpEqJ} are shown for a chain of length $N=12$, $S=6$, and $t/J_K = 0.1$. Colors correspond to $J = t/9N$ (blue), $J = t/6N$ (magenta, $J = t/4N$ (black), and $J=J_c = 0.029t/N$ (red). At the critical value $J=J_c$ the energy of the $p=\pi$ mode vanishes, indicating an instability of the ferromagnet.  }
    \label{fig:1D-LSWT-J}
\end{figure}

In general, Fig.~\ref{fig:LSWT-comp} shows that in this toy model problem the linear spin wave theory provides a good description of the small momentum excitations, but deviates significantly for the large momentum excitations around $p\approx \pi$. This deviation is strongly dependent on $S$ and becomes smaller as $S$ increases---as expected for a $1/S$ expansion.

Next, consider $1/S$ quantum corrections to the linear spin wave energies, as given by Eq.~\eqref{eq:E_q-SW}. In Fig.~\ref{fig:SWT-comp} we present a comparison between the exact magnetic excitation energies for a chain of length $N=32$ and the corresponding spin wave wave energies. The red and blue dots indicate the exact energies for $S=10$ and $S=30$, calculated using Eq.~\eqref{eq:ImpEqJ}, and the red and blue curves show the spin wave energies in the strong coupling limit (i.e., $t/J_K$ corrections neglected) up to order $1/S^2$. The dashed black curve shows the strong coupling linear spin wave result. We find that the $1/S$ corrections to linear spin wave theory reduce the spin wave energies, with the most pronounced effect occurring at large momenta $p\approx \pi$. This is an agreement with a previous result obtained in the context of a cubic lattice double-exchange model~\cite{Shannon:2002p104418}.

\subsection{Instability of the ferromagnet \label{ssec:FM-instability}}

\begin{figure}
    \centering
    \includegraphics[width=0.75\columnwidth]{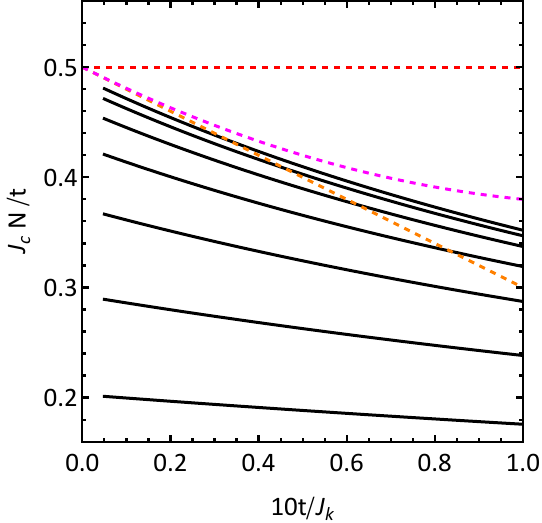}
    \caption{Critical Heisenberg coupling $J_c$ for which the ferromagnet becomes unstable, as a function of $t/J_K$. Black solid curves show the exact result for a chain of length $N=8$ and $S = 1, 2, 4, 8, 16, 32, 64$ (from bottom to top), obtained by determining when the energy of the $p=\pi$ magnetic excitation vanishes. The colored dashed curves represent predictions from linear spin wave theory. The red dashed line corresponds to the strong coupling limit; the dashed orange and magneta curves include order $t/J_K$ and $t^2/J_K^2$ corrections, respectively.}
    \label{fig:criticalJplot}
\end{figure}

We now turn to an analysis of the stability of the ferromagnetic ground state when an antiferromagnetic $J$ is included. An instability of the ferromagnet is expected since $J$ competes with the ferromagnetic tendency of the Kondo Hamiltonian. 

A basic understanding of this instability can be gained by considering the spin wave energies given by \eqref{eq:E_q-SW}. In the strong coupling limit, the linear spin wave energies are given by $E_p - E_0 \simeq ( t/N -2J) (1-c_p)/S$, which suggest an instability at a critical value $J_c = t/2N$. This does not reveal the nature of the instability, however, since the spin wave energies vanish for all $p$. Including the $t/J_K$ corrections to the linear spin wave energies reveals an instability at momentum $p=\pi$, marked by the appearance of a zero energy magnon (with respect to the ground state) at a critical value $J_c = (t-4t^2/J_K)/2N$. Note that the critical value of $J$ scales as $\sim 1/N$ due to the extreme dilute limit of a single electron. 

The exact value of the Heisenberg coupling $J_c$ at which the ferromagnet becomes unstable can be determined from Eq.~\eqref{eq:ImpEqJ} by tracking the energies of the magnetic excitations as $J$ is increased. This is shown in Fig.~\ref{fig:1D-LSWT-J} for a chain of length $N=12$ with $S=6$, where we have furthermore chosen $t/J_K = 0.1$. Figure \ref{fig:1D-LSWT-J} demonstrates that the energy of the $p=\pi$ magnetic excitation vanishes at a critical value $J_c$, in agreement with the spin wave theory result.

To further assess how well linear spin wave theory captures this instability at $p=\pi$, we show a comparison between linear spin wave theory and the exact result in Fig.~\ref{fig:criticalJplot}. Here, the exact value of $J_c$ is shown as a function of $t/J_K$ for different values of $S$, as well as the value of $J_c$ obtained from spin wave theory, represented by the dashed curves. The red curve corresponds to the strong coupling limit, which shows no dependence on $t/J_K$. The orange and magenta curves include first order and second order $t/J_K$ corrections, respectively. As expected, linear spin wave theory is in good agreement with the exact result for large $S$. It is furthermore clear from Fig.~\ref{fig:criticalJplot} that including $t/J_K$ corrections is important.


\section{Kondo chain in one dimension: spin polaron  \label{sec:1D-polaron}}

As discussed in Sec.~\ref{ssec:1D_exact}, the exact spectrum of the 1D Kondo chain, shown in Fig.~\ref{fig:KH_1D_exact}, features a band of spin polaron states, which can be understood as bound states of a spin-down electron and spin-flip excitations. This section is devoted to a more careful study of these spin polaron states. 

The special case of $S=1/2$ spins was considered in Ref.~\onlinecite{Sigrist:1991p2211}, where it was shown that the zero momentum ($p=0$) spin polaron solution is the exact ground state of the 1D Kondo chain when the Kondo coupling is antiferromagnetic (i.e., $J_K<0$ in our convention). Note that the analysis of Ref.~\onlinecite{Sigrist:1991p2211} did not include the Heisenberg interaction $J$. Here our goal is to consider the case of general $S$, examine the structure of spin polarons using the exact wave function, and compare the exact result to the result obtained from our canonical spin wave expansion. 

The main motivation to specifically focus on the spin polaron band is that it provides particularly useful insight into the nature of the spin wave expansion. In fact, examining the ground state of the 1D Kondo chain with antiferromagnetic Kondo coupling exposes one of the key features of the spin wave expansion, namely that electron operators in this spin wave formalism precisely correspond to the fermionic spin polaron excitations. This is a consequence of the strong coupling expansion via canonical transformation. The transformed electron operators, i.e., the electron operators after canonical transformation, create or annihilate electrons in a state of total spin $S-1/2$ with the local moment, and when expressed in the original basis, these states correspond to electrons dressed with spin wave excitations. Our goal in this section is to demonstrate this explicitly through the comparison of exact and spin wave theory results. 

For the purpose of this analysis is it useful to consider the case of antiferromagnetic Kondo coupling, such that the zero momentum spin polaron state is the ground state. This is readily accomplished by making the change $J_K \rightarrow - J_K$ in Eq.~\eqref{eq:1D-KH}, while keeping $J_K$ positive. This change is assumed in the remainder of this section. 

\begin{figure}
    \centering
    \includegraphics[width=\columnwidth]{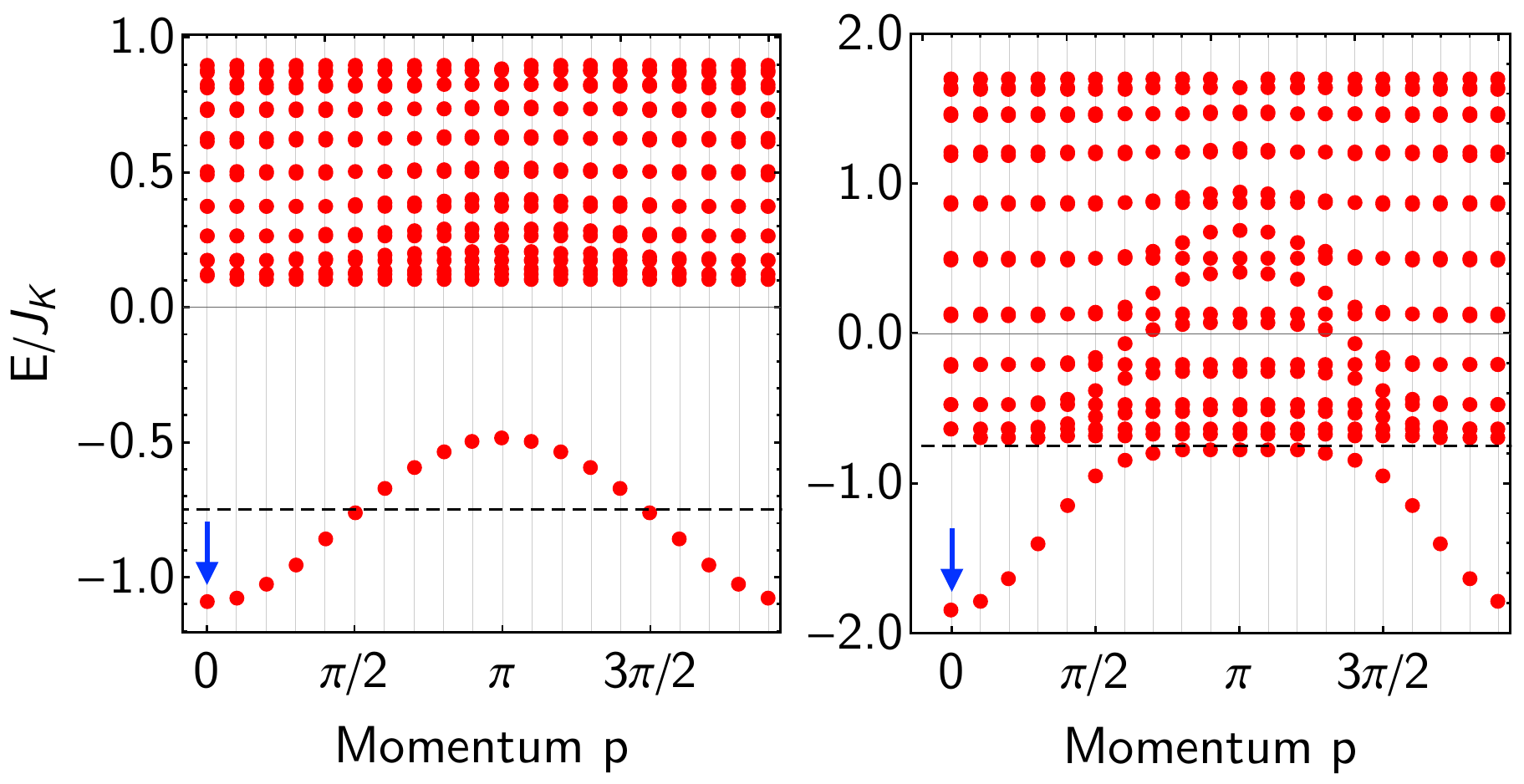}
    \caption{Exact spectrum of the 1D chain with antiferromagnetic Kondo coupling in the total spin $T=NS-1/2$ sector, as obtained from Eq.~\eqref{eq:ImpEqJ} after changing $J_K \rightarrow -J_K$ and setting $J=0$. Here we have chosen $N=20$ and $S=2$. The left and right panels show the spectrum for $t/J_K=0.2$ and $t/J_K=0.6$, respectively. The horizontal dashed line corresponds to $-J_K(1+1/S)/2$, i.e., the on-site Kondo energy of an electron forming a spin $S-1/2$ state with the local moment. Blue arrows indicate the ground state of Kondo chain with antiferromagnetic Kondo coupling.}
    \label{fig:polaron_spectrum}
\end{figure}

\subsection{Structure of the polaron: exact results \label{sec:polaron-exact}}

The exact spectrum in the $T=NS-1/2$ sector, calculated using Eq.~\eqref{eq:ImpEqJ}, is shown in Fig.~\ref{fig:polaron_spectrum}. Here we have chosen $N=20$ and $S=2$, and show the spectrum for two different values of the hopping: $t/J_K=0.2$ (left) and $t/J_K=0.6$ (right). Note that throughout this section we take $J=0$. 

\begin{figure}
    \centering
    \includegraphics[width=\columnwidth]{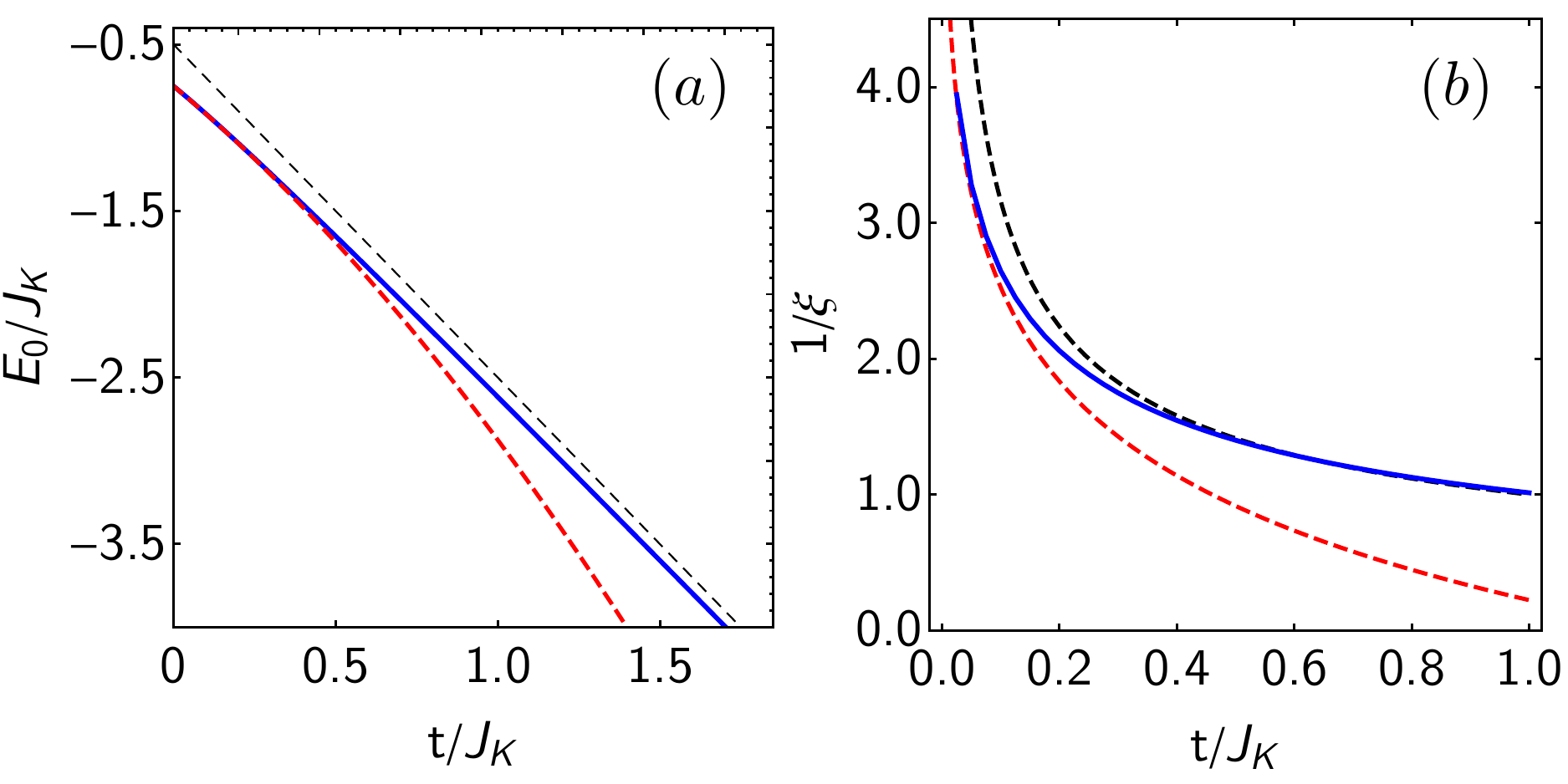}
    \caption{(a) Ground state energy $E_0$ of the zero momentum spin polaron as a function of $t/J_K$. Blue solid curve shows the exact result obtained from Eq.~\eqref{eq:ImpEqJ} for $N=20$ and $S=2$. The black dashed curve shows the classical energy $-J_K/2-2t$. The red dashed curve shows the spin wave expansion result given by Eq~\eqref{eq:E_0-polaron}. (b) Inverse polaron correlation length as a function of $t/J_K$. Blue curve corresponds to the exact result given by Eq.~\eqref{eq:xi-inverse} (calculated for $N=20$ and $S=2$). The red and black dashed curves show the approximate form of $\xi^{-1}$ when $t/J_K \rightarrow 0 $ and $t/J_K \gg1 $, respectively. }
    \label{fig:polaron_energy}
\end{figure}

Due to the antiferromagnetic Kondo coupling the spin polaron band in Fig.~\ref{fig:polaron_spectrum} is below the two-particle continuum. For any value of $t/J_K$ the ground state of the 1D Kondo chain is the zero momentum $p=0$ spin polaron state, indicated by blue arrows in Fig.~\ref{fig:polaron_spectrum}. As in the preceding sections, our focus is specifically on the regime of strong Kondo coupling. The energy of the ground state, denoted $E_0$, is shown as a function of $t/J_K$ in Fig.~\ref{fig:polaron_energy}(a). The black dashed line corresponds to the classical energy $-J_K/2-2t$ of a zero-momentum electron anti-aligned with ferromagnetically ordered classical local moments. The exact ground state energy, shown by the blue curve, is below the classical result, and the difference can be interpreted as the binding energy of the spin polaron. The binding energy decreases with increasing $S$ and the exact ground state energy will asymptotically approach the classical energy as $S\rightarrow \infty$ (i.e., the classical limit).

To examine the structure of the spin polaron, we follow Ref.~\onlinecite{Sigrist:1991p2211} and compute the correlation function
\be
C(r_i-r_j) =  N( \langle \bs^e_i \cdot \bS_j  \rangle - \langle \bs^e_i  \rangle \cdot  \langle \bS_j  \rangle ) /S, \label{eq:C-definition}
\ee
where $\bs^e_i $ is the electron spin and $ \bS_j $ is the local moment spin. Expectation values are computed in the ground state $\ket{\Psi}$, which is given by Eq.~\eqref{eq:Psi-def} with the appropriate ground state wave function components. In one dimension an exact closed form expression for the correlator can be obtained~\cite{Tsunetsugu:1997p809}. We find 
\be
C(r)=\frac{   1}{ \Lambda}\left[1 -
 \left( 1+ \frac{e^{-r/\xi}}{2S}\sqrt{\frac{J_K-2E_0 -4t }{J_K-2E_0 +4t }} \right)^2 \right], \label{eq:C-exact}
\ee
where $\Lambda$ is given by
\be
\Lambda = 1 + \frac{(J_K -2E_0-4t)^{1/2}( J_K -2E_0)}{2S( J_K -2E_0+4t)^{3/2}}, \label{eq:Lambda}
\ee
and the inverse correlation length is given by
\be
\xi^{-1} = \cosh^{-1}\left( \frac{J_K -2E_0}{4t}\right). \label{eq:xi-inverse}
\ee
Note that $r$ in \eqref{eq:C-exact} is measured in units of the lattice constant. 

The expression for $C(r)$ is valid for general $t/J_K$ and $S$. It is instructive, however, to consider two limiting cases: ({\it i}) the case $t/J_K\rightarrow 0 $ and ({\it ii})  the case $t/J_K\gg 1 $. In the first case, which corresponds to the strong coupling limit, we expect the electron to be correlated with essentially only one local moment and to form a spin $S-1/2$ state with the local moment spin $S$. Such a localized polaron has energy $E_0 \approx -J_K(S+1)/2S$ and the correlation function becomes
\be
C(r_i-r_j)\approx C_0 \delta_{ij}, \quad C_0 =  - \frac{4S+1}{2S(2S+1)}. \label{eq:C-strong}
\ee
The inverse correlation length takes the asymptotic form $\xi^{-1} \simeq \ln [(2S+1)J_K/2St]$. In the opposite weak coupling limit, when $t/J_K$ is large, the correlation function takes the approximate form
\be
C(r) \approx \frac{1}{1+(8S \xi)^{-1}}\left[1 - \left(1 + \frac{e^{-r/\xi}}{4S \xi } \right)^2 \right], \label{eq:C-weak}
\ee
with asymptotic correlation length $\xi \simeq \sqrt{t/J_K}$.

\begin{figure}
    \centering
    \includegraphics[width=0.7\columnwidth]{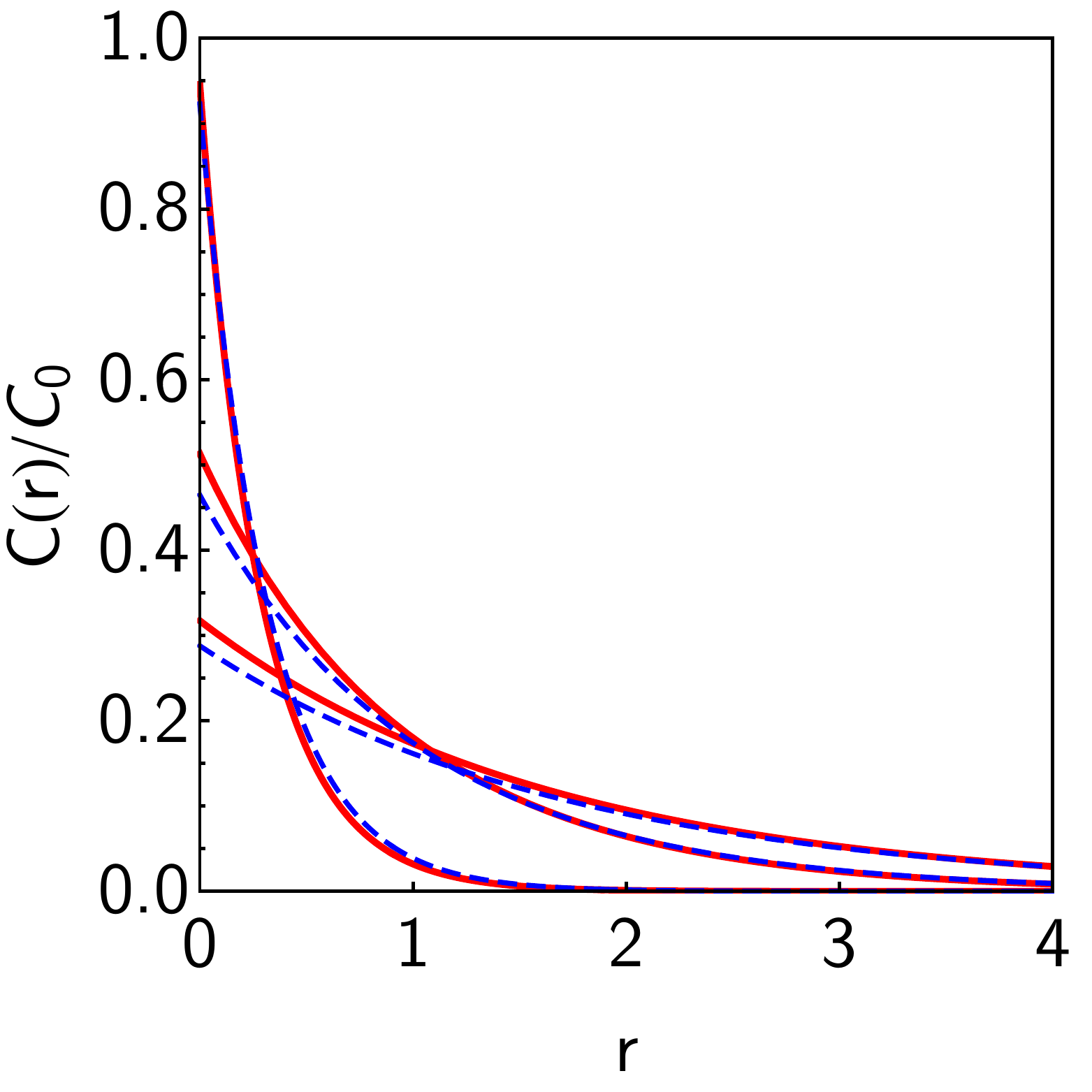}
    \caption{Polaron correlation function $C(r)$ as a function of distance $r$ (measured in units of lattice constant), as given by Eq.~\eqref{eq:C-exact}, normalized by $C_0$ defined in \eqref{eq:C-strong}. Red and blue dashed curves correspond to $S=2,8$, respectively. Three values of $t/J_K=(0.05,1.0,3.0)$ are plotted, shown from top to bottom (at $r=0$). }
    \label{fig:polaron_correlator}
\end{figure}

In Fig.~\ref{fig:polaron_energy}(b) we show the inverse correlation length as a function of $t/J_K$. The exact expression given by Eq.~\eqref{eq:xi-inverse} is compared with the asymptotic forms of the strong and weak coupling limits, which shows good agreement in the respective limits. In Fig.~\eqref{fig:polaron_correlator} we show the correlator $C(r)$ of Eq.~\eqref{eq:C-exact} as a function of $r$ for three different values of $t/J_K$, normalized by $C_0$ defined in \eqref{eq:C-strong}. The three values of $t/J_K$ are $(0.05,1.0,3.0)$. The red and dashed blue curves correspond to $S=2$ and $S=8$, respectively. It is clear from Fig.~\eqref{fig:polaron_correlator} that the size of the polaron increases as $t/J_K$ increases, and that the spatial extent of the polaron vanishes in the strong coupling limit, as expected from Eq.~\eqref{eq:C-strong}.

\subsection{Structure of the polaron: spin wave expansion \label{sec:polaron-expand}}

We now examine the structure of the spin polaron from the perspective of spin wave theory and demonstrate the central result of this section: that the electron operators of the strong coupling spin wave expansion are in precise correspondence with the spin polaron states. 

Within the framework of the strong coupling expansion the low-energy degrees of freedom of the Kondo chain with antiferromagnetic Kondo coupling are electrons in a state of spin $S-1/2$ with the local moments. We denote the corresponding operators as $d_i$. (See Appendix~\ref{app:SWT} for more details.) The ground state of the one-electron Kondo chain is then given by
\be \label{eq:polaronstate}
d_{0}^{\dagger}\ket{\text{FM}} = \frac{1}{\sqrt{N}}\sum_{i = 1}^{N} d_{i}^{\dagger}\ket{\text{FM}},
\ee 
which is a zero momentum superposition of electrons in a spin $S-1/2$ state. Our key finding, which we demonstrate below, is that the $d_i$ operators create and annihilate spin polarons. Hence, $d_{0}^{\dagger}$ creates a zero momentum spin polaron, in full agreement with the exact ground state discussed in Sec.~\ref{sec:polaron-exact}. This can be shown in a number of different ways. 

We first consider the energy of the state \eqref{eq:polaronstate} computed using the strong coupling spin wave Hamiltonian, the form of which is presented and discussed in more detail in Appendix~\ref{app:SWT} [see Eq.~\eqref{app:H-down}]. We find that this yields an energy given by
\begin{multline}
E_0 = -\frac{J_K}{2}  \left(1+ \frac{1}{S} \right) -2t \left(1- \frac{1}{2S}  +\frac{1}{4S^2}\right) \\
 - \frac{t^2}{J_K} \left( \frac{3}{S}  - \frac{8}{2S^2}\right) + \ldots. \label{eq:E_0-polaron}
\end{multline}
The energy is shown as a function of $t/J_K$ in Fig.~\ref{fig:polaron_energy} (dashed red curve), showing very good agreement with the exact solution for small values of $ t/J_K \lesssim 0.5 $. The agreement with the exact solution can be made more precise by expanding Eq.~\eqref{eq:ImpEqJ} in $1/S$ and $t/J_K$ on both sides and solving for $E_0$ order-by-order in the expansion parameters, which is equivalent to an expansion of the exact ground state energy $E_0$ in $1/S$ and $t/J_K$. Comparing the result with \eqref{eq:E_0-polaron} computed within spin wave theory we find perfect agreement. 

Next, we consider the structure of the operators $d_i$ in terms of the original electron operators $c_{i\sigma}$ and the local moment spin flip operators $S^+_i \simeq \sqrt{2S} a_i$. The spin wave expansion defines a transformation relating the eigenstates of the Kondo coupling (given by $d_i$) and the constituent electron and local moment spins, which can be used to express one in terms of the other as an expansion $1/S$ and $t/J_K$. This expression is provided in \eqref{app:d-fermion} of Appendix~\ref{app:SWT} and suggests that the $d_i$ operators should indeed be viewed as $c_{i\down}$ operators dressed with spin flip excitations, as expected for a spin polaron.

\begin{figure}
    \centering
        \includegraphics[width=\columnwidth]{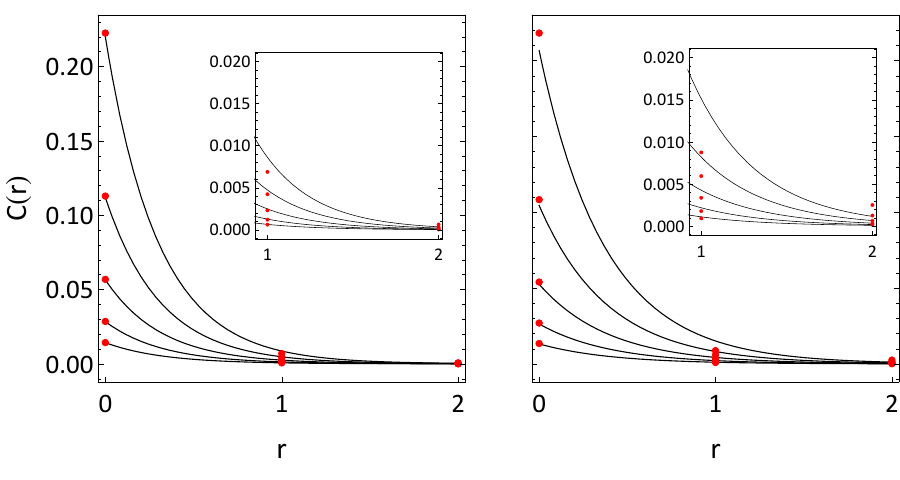}
    \caption{Comparison of the exact expression for the correlator $C(r)$, as given by Eq.~\eqref{eq:C-exact}, with the result from spin wave theory, as given by Eq.~\eqref{eq:C(r)-SWT}, for different values of $S$. Black curves correspond to exact expression and red dots correspond to $C(r)$ of Eq.~\eqref{eq:C(r)-SWT}. (a) $t/J_K = 0.05$ and $S = 4, 8, 16, 32, 64$ (top to bottom). (b) Same as $(a)$ but for $t/J_K = 0.1$. Insets show enlarged view of $r = 1, 2$.}
    \label{fig:C(r)_compare}
\end{figure}

To confirm this, we compute the correlation function $C(r)$ defined in Eq.~\eqref{eq:C-definition} within the framework of spin wave theory. This is achieved by substituting the local moment spins in Eq.~\eqref{eq:C-definition} with Holstein-Primakoff bosons, i.e., by setting $S^z_i = S- a^\dagger_i a_i$ and $S^+_i = (S^-_i )^\dagger \simeq \sqrt{2S} a_i$, and calculating averages with respect to the ground state \eqref{eq:polaronstate} after expressing all $d_i$ operators in terms of $c_{i\sigma}$ and $a_i$. Details are presented in Appendix~\ref{app:polaron}. In this way, we are able to determine $C(r)$ for $r=0,1,2$ as an expansion in $1/S$ and $t/J_K$, and find
\be
\begin{split}
C(0) &= - \frac{1}{S} + \frac{1}{4S^2} + \frac{2t}{J_KS} - \frac{6t^2}{J_K^2S} -\frac{3t}{J_KS^2}, \\
C(1) & = -\frac{t}{J_KS} + \frac{4t^2}{J_K^2S} + \frac{t}{J_KS^2},\\
C(2) & =  -\frac{t^2}{J_K^2S}.
\end{split}   \label{eq:C(r)-SWT}
\ee
These values for the correlator at distances $r=0,1,2$ can be compared to the exact result of Eq.~\eqref{eq:C-exact}. In Fig.~\ref{fig:C(r)_compare} we show the exact correlator $C(r)$ for different values of $S$ and indicate the approximate result from spin wave theory, given by \eqref{eq:C(r)-SWT}, by red dots. This comparison confirms the expected behavior: the agreement becomes better as $t/J_K$ and $1/S$ become smaller.  

To further examine the agreement with the exact result we can go one step further and expand the exact correlator $C(r)$ as given by Eq.~\eqref{eq:C-exact} in small $t/J_K$ and $1/S$ for $r=0,1,2$. Expansing to the same order as in \eqref{eq:C(r)-SWT} we find perfect order-by-order agreement. This demonstrates that the ground state of Eq.~\eqref{eq:polaronstate} indeed describes a spin polaron.

\section{Discussion and Conclusion \label{sec:discuss}}

In this work we have examined a class of Kondo-Heisenberg toy models with quantum spins for which the ground state and the lowest energy excitations can be established exactly. The class of toy models is defined by $N$ spins forming a 1D chain and a single electron Kondo coupled to the spins. Our central result is a detailed analysis of the energies and eigenstates of the magnetic excitations of the ferromagnetic ground state, and in particular a comparison between the exact solutions and the approximate result obtained from a strong coupling spin wave expansion~\cite{our_paper}. We find that spin wave theory correctly captures all aspects of the full quantum problem, i.e., the energy and structure of the eigenstates, order-by-order in $1/S$ and $t/J_K$, the perturbative parameter of the strong coupling regime. 

The case of the 1D Kondo chain with antiferromagnetic Kondo coupling is of particular interest, since the exact ground state is a ferromagnet in which the electron forms a bound state with spin flip excitations---a spin polaron. This implies that even the ground state already has nontrivial quantum structure not captured by a classical Kondo lattice model. Our second key result is that the electron operators of the canonical spin wave expansion, which are related to the bare electron operators by canonical transformation and describe electrons in a state of spin $S-1/2$ with the local moments, precisely correspond to spin polaron of the exact ground state. This result is demonstrated by computing the correlation function of electron and local moment spins, which captures the structure of the spin polaron, both exactly and within the spin wave expansion scheme. The found order-by-order agreement is a direct consequence of the nontrivial structure of the electron operators in the strong coupling canonical spin wave expansion. 

In a broader sense, our work gives insight into the general accuracy of spin wave expansions in itinerant magnets described by exchange coupled electrons and spins. By comparing with exactly solvable toy models, our work in particular provides a better understanding of how well $1/S$ expansions capture the magnetic excitations in itinerant magnets, more specifically itinerant ferromagnets. Evidently, even considering the case of a single electron coupled to local moment spins highlights how the quantum structure of itinerant ferromagnets is fundamentally different from Heisenberg ferromagnets. In the latter, the spin wave excitations are known to be exact eigenstates of the Hamiltonian, and are thus in precise agreement with the exact magnetic excitations. This is not the case in itinerant magnets, as the toy models discussed in this paper powerfully demonstrate. Our work shows how, within the confines of these toy models, the energy of magnetic excitation depends on $1/S$ and how well spin wave theory captures these excitations. Our analysis furthermore reveals how well spin wave theory captures the instability of the ferromagnet.

\mbox{}

\section*{Acknowledgements}

We have greatly benefited from conversations and correspondence with Urban Seifert, Alexander Chernyshev, Rafael Fernandes, Pok Man Tam, Sanjeev Kumar, and Maria Daghofer. This research was supported by the National Science Foundation Award No. DMR-2144352.

\appendix

\section{Diagonalization of the KH dimer \label{app:dimer}}

Consider a single site with a local spin $S$ and one electron. It is straightforward to construct states of total spin using the quantum theory of angular momentum. The total spin can take values $S\pm 1/2$ and the corresponding eigenstates are given by
\beq
\ket{S+\tfrac12,M} &=& u \ket{S,M-\tfrac12; \up} + v \ket{S,M+\tfrac12; \down}, \label{S+1/2} \\
\ket{S-\tfrac12,M} &=& -v \ket{S,M-\tfrac12; \up} + u \ket{S,M+\tfrac12; \down},  \label{S-1/2} 
\eeq
where $u=u_M$ and $v=v_M$ are the appropriate Clebsch-Gordan coefficients given by 
\be
u_M = \sqrt{\frac{S+1/2+M}{2S+1}}, \quad v_M = \sqrt{\frac{S+1/2-M}{2S+1}}.
\ee
These states are eigenstates of the local Kondo coupling $-J_K \bS\cdot \bs/S$ with energies $-J_K/2$ (for total spin $S+1/2$) and $-J_K(S+1)/2S$ (for total spin $S-1/2$). It is worth pointing out that $\ket{S+\tfrac12,S+\tfrac12} = \ket{S,S; \up} $ is a simple product state, whereas
\be
\ket{S-\tfrac12,S-\tfrac12} =\frac{1}{\sqrt{2S+1}} ( \sqrt{2S} \ket{S,S; \down}-\ket{S,S-1; \up} ),
\ee
is not. It follows that the states $\ket{S-\tfrac12,M} $ are never product states and there is thus a fundamental distinction between an electron and a spin in a quantum state of total spin $S\pm 1/2$.

For the dimer problem, when we consider two sites and one electron, the states of total spin $T$ are given by Eqs.~\eqref{states-1} and \eqref{states-2}. The expressions for these total spin states feature the Clebsch-Gordan coefficients 
\be
C^{TM}_{S\pm\frac12,M_1;SM_2} ,\;C^{TM}_{SM_1;S\pm\frac12,M_2} ,
\ee
reflecting the fact that, at the site where the electron sits, it forms a state of total spin $S\pm 1/2$ with the local moment spin. In general, Clebsch-Gordan coefficients have the symmetry property 
\be
C^{JM}_{j_1 m_1;j_2m_2}  = (-1)^{j_1+j_2-J} C^{JM}_{j_2m_2;j_1 m_1} ,\label{app:CG-sym}
\ee
and this symmetry property plays an important role for determining the eigenvalue under site-exchange (i.e., inversion) of the states given by~\eqref{states-symmetrized}. Indeed, when applied to the present case we have $j_1 = S\pm 1/2$, $j_2=S$, and $J=T$, from which it follows that $(-1)^{j_1+j_2-J}  = (-1)^{2S\pm 1/2-T} $. It is this symmetry property of the Clebsch-Gordan coefficients which explains the eigenvalue under inversion quoted in Sec.~\ref{ssec:dimer} of the main text.

\section{Classical limit of KH dimer \label{app:dimer-classical}}

In this Appendix we briefly review the classical limit of the KH dimer. In the classical limit, the spins are treated as classical vectors and the Heisenberg energy is simply $J\cos \theta$, where $\theta$ is the angle between the two spins of the dimer. To determine the electronic energy as a function of $\theta$, we solve the electronic Hamiltonian. This is straightforward in the classical limit since it is quadratic (i.e., non-interacting). Expressed in a site-dependent spin basis tied to the direction of the local moments, the Hamiltonian takes the form
\be
H_e = \begin{pmatrix}  A & B \\ B^\dagger & A\end{pmatrix} ,
\ee
where $A$ and $B$ are given by
\be
A = -\frac12 J_K  \begin{pmatrix}  1 & 0 \\ 0 & -1\end{pmatrix} , \quad B= -t \begin{pmatrix}  \cos\frac{\theta}{2} & -\sin\frac{\theta}{2}  \\ \sin\frac{\theta}{2}  & \cos\frac{\theta}{2} \end{pmatrix} .
\ee
Here $A$ describes the Kondo coupling, which is diagonal in the chosen local frame, and $B$ describes the (spin-dependent) hopping between the sites. 

It is straightforward to diagonalize the electronic Hamiltonian. It is sufficient to consider $J_K>0$ and focus on the lower two spectral branches, since the upper spectral branches are obtained by setting $J_K \rightarrow -J_K$. Expanding the energies up to second order in $t/J_K$, lower two branches of the KH dimer are given by
\be
E^c_\pm \approx -\frac12 J_K  \pm t  \cos\frac{\theta}{2} - \frac{t^2}{J_K} \sin^2\frac{\theta}{2} + J\cos\theta,
\ee
which may be rewritten as
\begin{multline}
E^c_\pm \approx -\frac12 J_K  - \frac{t^2}{J_K}-J\pm t  \cos\frac{\theta}{2} \\
+ \left(2J + \frac{t^2}{J_K}\right) \cos^2\frac{\theta}{2} .
\end{multline}
This is the expression for the energies used in Fig.~\ref{fig:dimer_afm}. 

It is instructive to examine the two special cases $\theta=0$ and $\theta=\pi$, i.e., the classical ferromagnet and anti-ferromagnet, respectively. The classical ferromagnet has energy $- J_K/2  + J\pm t $, which coincides with the exact solution of the quantum ferromagnet, i.e., the state with total spin $T=2S+1/2$.  The classical anti-ferromagnet has energy 
\be
E^{c,\text{AF}}_\pm \approx -\frac12 J_K  - \frac{t^2}{J_K}-J,
\ee
which is the energy indicated by orange dashed in lines in Fig.~\ref{fig:dimer_exact}.

\section{Canonical spin wave theory \label{app:SWT}}

This Appendix presents further details on the strong coupling spin wave expansion, in particular the full form of the Hamiltonian describing the low-energy charge and magnetic excitations. The spin wave expansion scheme used in this work was developed by the present authors in a separate work~\cite{our_paper}; for details beyond what is collected in this Appendix we refer the reader to Ref.~\onlinecite{our_paper}. 

In general, spin wave theories are expansions around a classical ground state and describe the low-energy excitations above that ground state. Here, in the regime of strong Kondo coupling, when $J_K$ is the largest energy scale, the proper low-energy degrees of freedom are electrons in a state of total spin $S\pm 1$ (i.e., eigenstates of the on-site Kondo coupling) and spin flip excitations of that total spin at each site. The state of the low-energy electrons depends on the sign of the Kondo coupling: when $J_K>0$ ($J_K>0$) they form a state of total spin $S+1/2$ ($S-1/2$) with the local moment. The proper magnons are fluctuations of that total spin. A key feature of the spin wave expansion is that it provides an expression for the proper degrees of freedom, which correspond to the total spin in a given site, in terms of the bare degrees of freedom, which correspond to the constituent local moment and electron spin. The transformation between these degrees of freedom is determined as an expansion in $1/S$, as well as $t/J_K$ and $J/J_K$. See Eqs.~\eqref{app:f-fermion}, \eqref{app:a-boson} and \eqref{app:d-fermion} below.

{\it The case of positive $J_K$ (FM Kondo coupling).} Consider first the case where $J_K$ is positive. The fermion operators corresponding to the electrons in a state of total spin $S+1/2$ are denoted $f_i$ and the spin flip excitations are described by boson operators $a_i$. The spin wave Hamiltonian governing the dynamics of these excitations is an expansion in $1/S$ (as usual for a spin wave expansion) and in the small parameters $t/J_K$, $J/J_K$. Up to order $1/S^2$, keeping all perturbative corrections of the form $t^n J^m/J^{n+m}_{K}$ with $n+m \le 2$, the Hamiltonian for $f_i$ and $a_i$ is given by
\begin{widetext}
\begin{multline}
\mathcal H = \sum_{ij} t_{ij} f^\dagger_{i}f_{j}+ \frac{1}{2S}\left(1-\frac{1}{2S}\right) \sum_{ij} t_{ij} a^\dagger_ia_j f^\dagger_{i}f_{j}-\frac{1}{4S}\left(1-\frac{3}{8S}\right) \sum_{ij} t_{ij} \left( a^\dagger_ia_i+a^\dagger_ja_j\right) f^\dagger_{i}f_{j}\\
-\frac{1}{4J_K S}  \sum_{ijk} t_{ik} t_{kj} \left(2  a^\dagger_i a_j + a^\dagger_ia_i +a^\dagger_j a_j- 2 a^\dagger_ka_j- 2 a^\dagger_ia_k  \right)  f^\dagger_{i}f_{j} + \mathcal H_J \label{app:H-up}
 \end{multline}
 \end{widetext}
where $\mathcal H_J$ collects all terms coming from the Heisenberg coupling:
\begin{multline}
\mathcal H_J =  \frac{1}{2S}\sum_{ij}J_{ij}(a_i^{\dagger}a_j - a_i^{\dagger}a_i -a_j^{\dagger}a_j ) \\
+\frac{1}{4S^2}\sum_{ij}J_{ij}(2a_i^{\dagger}a_i  - a_i^{\dagger}a_j -a_j^{\dagger}a_i ) f^\dagger_{i}f_{i} \\
-\frac{1}{2J_KS^2}\sum_{ijk}t_{ij}\left[J_{ik}(a_i^{\dagger}a_k-a_k^{\dagger}a_j+a_i^{\dagger}a_j - a_i^{\dagger}a_i)+\right. \\
\left.   J_{jk}(a_k^{\dagger}a_j-a_i^{\dagger}a_k+a_i^{\dagger}a_j-a_j^{\dagger}a_j) \right] f_{i}^{\dagger}f_{j}. \label{app:H_J-up}
 \end{multline}
It is important to note that the form of the spin wave Hamiltonian, as is usual, depends on the classical magnetic ground state. Here we have performed an expansion around the ferromagnet, which is the subject of this work. 

When applied to the two-site dimer, and after expressing the Hamiltonian in terms of bonding and anti-bonding operators defined in Eq.~\eqref{eq:0-pi-operators}, we obtain Eqs.~\eqref{eq:H_0} and \eqref{eq:H_1}.

When applied to the 1D KH chain, we obtain the (Fourier transformed) spin wave Hamiltonian 
\begin{multline}
\mathcal H \simeq \sum_p (-\tfrac12 J_K-2tc_p ) f^\dagger_p f_p + \sum_q \Omega_q a^\dagger_q a_q f^\dagger_0 f_0 \\
+ \sum_{q,p} \Gamma_{q,p} a^\dagger_p f^\dagger_{q-p} a_q  f_0 + \mathcal H_J, \label{eq:H-SW-1D}
\end{multline}
where $\Omega_q$ and $\Gamma_{q,p} $ are given by
\begin{align}
\Omega_q & =\frac{t}{SN}\left[1-c_q -\frac{2t(1-c_q)^2}{J_K} 
 +\frac{4c_q -3}{8S} \right] \\
\Gamma_{q,p} & = \frac{ t }{2SN}(1+c_{p-q}-2c_q )
\end{align}
and $\mathcal H_J $ takes the form
\begin{multline}
\mathcal H_J =   -\frac{2J}{S}\sum_q  \left(1-c_q\right)a^\dagger_q a_q f^\dagger_0 f_0 \\
+\frac{J}{S^2N}\sum_q \left[1-c_q - \frac{4t(1-c_q)^2}{J_K} \right] a^\dagger_q a_q f^\dagger_0 f_0.
\end{multline}
Note that in Eq.~\eqref{eq:H-SW-1D} we have dropped all terms from the more general Hamiltonian \eqref{app:H-up} which are unimportant for determining the energies of the spin wave excitations $ a^\dagger_q  f^\dagger_{0} \ket{\text{FM}}$, see Sec.~\ref{ssec:spin-waves}. Note further that the Hamiltonian of Eq.~\eqref{eq:H-SW-1D} is almost diagonal (i.e., $ a^\dagger_q  f^\dagger_{0} \ket{\text{FM}}$ are almost eigenstates) except for the scattering term with vertex $\Gamma_{q,p}$. Treating the scattering term to second order in perturbation theory, which yields contributions of the order $t/S^2$, gives the energies of Eq.~\eqref{eq:E_q-SW}.

The Hamiltonian of Eq.~\eqref{app:H-up} describes the dynamics of the low-energy charge and magnetic excitations. To understand the structure of these excitations in terms of the bare degrees of freedom (i.e., the electron operators $c_{i\sigma}$ and the local moment spin flips $S^\pm _i$), we expand the operators $f_i$ and $a_i$ in terms of $c_{i\sigma}$ and bosons corresponding to $S^\pm _i$. In the same way as the Hamiltonian $\mathcal H$, this expansion depends on the magnetic ground state. For a ferromagnet we find 
\begin{multline}
f_{i} \rightarrow     c_{i\up} + \frac{1}{\sqrt{2S}} \Big[ a_i^{\dagger}c_{i\down} +\sum_{j} \frac{t_{ij}}{J_K}(a_j^{\dagger} - a_i^{\dagger})c_{j\down}\Big] \\
 + \frac{1}{4J_KS}\sum_{j}t_{ij}(2a_i^{\dagger}a_j - a_i^{\dagger}a_i - a_j^{\dagger}a_j)c_{j\up} \label{app:f-fermion}
\end{multline}
and
\begin{multline}
a_i \rightarrow     a_i + \frac{1}{\sqrt{2S}} \Big[s_i^{+} +  \sum_j \frac{t_{ij}}{ J_K} \left( c_{j \uparrow}^{\dagger} c_{i \downarrow} -c_{i \uparrow}^{\dagger} c_{j \downarrow}\right) \Big]\\
  - \frac{1}{4S} \Big[ a_i c_{i\up}^{\dagger}c_{i\up}   
+ \sum_j \frac{t_{ij}}{J_K} (2a_j - a_i)c_{i\up}^{\dagger}c_{j\up} \Big] \label{app:a-boson}
\end{multline}
In both expressions, \eqref{app:f-fermion} and \eqref{app:a-boson}, the boson operators on the right hand side should be interpreted as $S^+ _i \simeq \sqrt{2S}a_i$, i.e., spin flips of the local moment spin $S$. Using these expressions we are able to establish \eqref{eq:E_pi_0-state} of the main text.

{\it The case of positive $J_K$ (AFM Kondo coupling).} Consider the next the case where $J_K$ is negative, implying an antiferromagnetic Kondo coupling. In this case the local moment spin and the electron want to form a $S-1/2$ state. Such a state is very different from a $S+1/2$ state. The fermion operators corresponding to the electrons in a state of total spin $S-1/2$ are denoted $d_i$; the spin flip excitations are still denoted $a_i$.

A key difference with the case previously considered is that here we are only interested in the dynamics of the $d_i$ electrons. Precisely because of the special nature of the electron states these operators correspond to, the low-energy Hamiltonian for the $d_i$ electrons already describes all the physics of interest. The relevant spin wave Hamiltonian $\mathcal H$ takes the form 
\begin{multline}
\mathcal{H} =  \left(1 - \frac{1}{2S} + \frac{1}{4S^2} \right)\sum_{ij}t_{ij}d_{i}^{\dagger}d_{j} \\ 
+ \frac{1}{2J_K S}\sum_{ijk}t_{ik}t_{kj}(1 + \delta_{ij})d_{i}^{\dagger}d_{j}+  \mathcal H_J \label{app:H-down}
\end{multline}
where $\mathcal H_J$ describes all contributions from the Heisenberg coupling and is given by
\begin{multline} 
\mathcal{H}_J =  -\frac{1}{2S^2} \sum_{ij}J_{ij}d_{i}^{\dagger}d_{i}  +\frac{1}{J_K S^2}\sum_{ij}t_{ij}J_{ij}d_i^{\dagger}d_j \\
+ \frac{1}{2J_KS^2}\sum_{ijk}t_{ij}(J_{ik} + J_{jk})d_i^{\dagger}d_j. \label{app:H_J-down}
\end{multline}
While $\mathcal H$ does not depend on the bosons $a_i$, it is manifestly a spin wave Hamiltonian, since it is an expansion in $1/S$. The form of $\mathcal H$ encodes the special nature of the $d_i$ fermions in the strong Kondo coupling regime. It is worth emphasizing that even though the original Hamiltonian describes an interaction between the local moments, its contribution to $\mathcal H$ given by \eqref{app:H_J-down} does not depend on the bosons but only on the fermions. 

The Hamiltonian of Eq.~\eqref{app:H-down} is used to compute the energy of the polaron ground state \eqref{eq:polaronstate}, which yields \eqref{eq:E_0-polaron}.

The form of \eqref{app:H_J-down} is a direct consequence of the structure of $d_i$ operators. As for the fermion operators $f_i$, see Eq.~\eqref{app:f-fermion}, the spin wave expansions yields and expression for the $d_i$ operators (which describe electrons in a quantum mechanical $S-1/2$ state) in terms of bare electron and spin flip operators. Specifically, one finds that
\begin{multline}
d_i \rightarrow    \left(1 - \frac{1}{4S} \right)c_{i\down} -  \frac{1}{\sqrt{2S}}\Big[ a_i c_{i\up} +\sum_{j} \frac{t_{ij}}{J_K}  (a_i - a_j)c_{j\up} \\
+\sum_{jk}\frac{t_{ik}t_{kj}}{J^2_K}(2a_k - a_j - a_i)c_{j\up}  \Big]  \\  
+ \frac{1}{4S\sqrt{2S}}\Big[ a_i c_{i\up} +\sum_{j} \frac{t_{ij}}{J_K}  (3a_i -5 a_j)c_{j\up}\Big] .\label{app:d-fermion}
\end{multline}
It is this form which clearly exposes the $d_i$ operators as creation and annihilation operators of spin polaron states. We use Eq.~\eqref{app:d-fermion} together with Eq.~\eqref{eq:polaronstate} to compute the polaron correlation function within spin wave theory. See Sec.~\ref{sec:polaron-expand} and Appendix~\ref{app:polaron}.

\section{Group theory analysis 1D KH chain \label{app:1D-group-theory}}

This Appendix provides a group theory analysis of the 1D KH chain. The goal is to determine the structure of the energy spectrum in terms of spatial symmetry quantum numbers. Such analysis is separate from, and complements, an analysis of the spin rotational symmetry of the problem, which simply implies that total spin is a good quantum number. 

We first examine the relevant symmetries and symmetry quantum numbers of the 1D chain. The 1D chain problem is defined by $N$ sites, with quantum spins of length $S$ located at each site and one itinerant electron on the entire chain. This system has translation symmetry and inversion symmetry. The inversion symmetry is denoted $\sigma$ and translations by a distance $R$ are denoted $t_R$. Since $R = n a$, where $a$ is the lattice constant and $n$ is an integer, one has $t_R= t_{na} = t^n_a$. Here $t_a$ is the translation by one lattice constant, i.e., the generator of the translation (sub)group. The requirement $t^N_a=1$ is the quantization condition for momentum; the (center-of-mass) momentum is a good quantum number. At inversion symmetric momenta all eigenstates can additionally be labeled by their inversion eigenvalues. (The full set of good quantum numbers includes total spin $T$ and magnetic quantum number $M$, the eigenvalue of $T^z$). 

To understand the structure of the spectrum using group theory, consider first the case where the electron is fixed at a particular site of the chain. At that site the electron forms a spin of total length $S' = S\pm 1/2$. Two spatial symmetries are then present: the identity $\mathbb{1}$ and the inversion $\sigma$, forming the group $\{\mathbb{1},\sigma\}$. The number of inversion centers depends on whether $N$ is even or odd. When $N$ is even there two inversion centers: the site where the electron sits and the site separated by $t^{N/2}_a$. Instead, when $N$ is odd only the site where the electron sits is an inversion center. 

States are either even or odd under $\sigma$, giving rise to two representations labeled $\pm$. In general, the number of times an irreducible representation $\Gamma$ is contained in a reducible representation of a group $G$ (here: the reducible representation of $G$ within the invariant subspace labeled by $T$) is given by
\be
n^T_{\Gamma} =  \frac{1}{h_G} \sum_{\mathcal C} \text{Tr}\,(D^T_{\mathcal C})\, \chi^\Gamma_{\mathcal C}\, N_{\mathcal C},
\ee
where $h_G$ is the order of the group, $\chi^\Gamma_{\mathcal C}$ is the character of the class $\mathcal C$, and $N_{\mathcal C}$ is the number of elements in $\mathcal C$. In the present case, this general group theoretical formula greatly simplifies:
\be
n^T_{\pm } =   \frac{1}{2}\text{Tr}\,(D^T_{\mathbb{1}})\pm  \frac{1}{2}\text{Tr}\,(D^T_{\sigma}).
\ee
To determine the trace of the reducible representations $D^T_{g}$ with $g\in \{\mathbb{1},\sigma\}$ we make use of the formula~\cite{Lecheminant:1995p6647}
\be
\text{Tr}\,(D^T_g) = \text{Tr}\,(D^{M=T}_g) -\text{Tr}\,(D^{M=T+1}_g).
\ee
Representations within a subspace of given $M$ are easier to construct than representations for given $T$; in particular, the traces are easy to compute. For the identity $\mathbb{1}$ we find the formula
\be
\text{Tr}\,(D^M_{\mathbb{1}}) = \sum_{i_1,\ldots,i_{N-1}=-S}^S\sum^{S'}_{i_N=-S'} \delta_{\sum^N_n i_n,M}.
\ee
The trace formula for the inversion $\sigma$ depends on whether $N$ is even or odd. If $N$ is even one has
\begin{multline}
\text{Tr}\,(D^M_{\sigma}) =\\
 \sum_{i_1,\ldots,i_{N/2-1}=-S}^S\sum_{i_{N/2}=-S}^S \sum_{j=-S'}^{S'} \delta_{2\sum^{N/2-1}_n i_n+i_{N/2}+j,M}.
\end{multline}
Instead, when $N$ is odd the trace is computed via
\be
\text{Tr}\,(D^M_{\sigma}) = \sum_{i_1,\ldots,i_{(N-1)/2}=-S}^S \sum_{j=-S'}^{S'} \delta_{2\sum^{(N-1)/2}_n i_n+j,M}.
\ee

We can now apply these formulas to the total spin sectors of primary interest, $T=NS+1/2$ and $T=NS-1/2$. Smaller values of total spin can be considered too, but are not of immediate relevance here. Consider first the case $T=NS+1/2$. The only contributions can come from $S'=S+1/2$ and one straightforwardly finds that $(n_+,n_-) = (1,0)$.

Now consider the case $T=NS-1/2$, and first assume $S'=S+1/2$. We have to distinguish even and odd $N$. In both cases one has $\text{Tr}\,(D^T_{\mathbb{1}})=N-1$. When $N$ is even one has $\text{Tr}\,(D^T_{\mathbb{1}})=1$ and when $N$ is odd one has $\text{Tr}\,(D^T_{\mathbb{1}})=1$. This implies that $(n_+,n_-) = (N/2,N/2-1)$ for $N$ even and $(n_+,n_-) = (N/2-1/2,N/2-1/2)$ for $N$ odd. 

Next, consider the case $S'=S-1/2$. This case is again trivial and one finds $(n_+,n_-) = (1,0)$ for both even and odd $N$.

\section{Exact solution 1D chain \label{app:1DJ}} 

This Appendix provides additional details of the solution to the 1D KH chain in the $T= NS-1/2$ sector. 

Consider the state $\ket{\Psi}$ in the $T^z= NS-1/2$ subspace given by Eq.~\eqref{eq:Psi-def}. It is useful to define $\ket{\Psi} \equiv  X \ket{\text{FM}}$, with the operator $X$ given by
\be
X = \sum_{i}\left(\Psi^i c^\dagger_{i\down} + \frac{1}{\sqrt{2S}}\sum_j\Psi^i_jc^\dagger_{i\up}S^-_j \right) .
\ee
Letting the Hamiltonian $H$ act on $\ket{\Psi}$ gives
\be
H \ket{\Psi} = H X \ket{\text{FM}} = [H,X]\ket{\text{FM}}+E^J_c\ket{\Psi},
\ee
where $E^J_c$ is the classical energy of a Heisenberg ferromagnet defined in Eq.~\eqref{eps_p}, and we have used $H\ket{\text{FM}}=E^J_c\ket{\text{FM}}$. The Schr\"odinger equation then takes the form 
\be
( H - E^J_c)\ket{\Psi} = [H,X]\ket{\text{FM}} = E\ket{\Psi} ,
\ee
where we have defined the energy $E$ relative to the constant classical energy $E^J_c$. Evaluating the commutator $[H,X]$ yields the set of equations given by Eqs.~\eqref{eq:Psi_i} and \eqref{eq:Psi_ij}.

In order to project out solutions in the sector of total spin $T= NS+1/2$ and retain only those solutions with total spin $T=NS-1/2$, we impose the constraint $T^+ \ket{\Psi} =0$. As described in the main text, $T^+ = T^x+iT^y= \sum_i s^+_i + S^+_i$ is the raising operator of total spin and the constraint can be rewritten as
\be
T^+ \ket{\Psi} =T^+X \ket{\text{FM}} = [T^+, X ] \ket{\text{FM}} =0,
\ee
Evaluating the commutator $ [T^+, X ] $ gives rise to a condition on the wave functions $\Psi^i$ and $\Psi^i_j$ given by Eq.~\eqref{constraint}. After Fourier transformation this constraint takes the form
\be
\widetilde\Psi^p + \sqrt{2SN}\widetilde\Psi^p_0 = 0. \label{app:constraint}
\ee
The Fourier transformed Schr\"odinger equation of Eqs.~\eqref{eq:Psi_p} and \eqref{eq:Psi_pq} must be solved subject to this constraint. 

The coupled set of equations \eqref{eq:Psi_p} and \eqref{eq:Psi_pq} can be solved by first using \eqref{eq:Psi_p} to express $\widetilde\Sigma^p$ in terms of $\widetilde\Psi^p$, and the substituting this expression into \eqref{eq:Psi_pq}. This yields Eq.~\eqref{eq:PsiPsi}, which automatically satisfies \eqref{app:constraint}. Summing over $q$ gives $\widetilde\Sigma^p$, which is then substituted back into \eqref{eq:Psi_p}. Equation \eqref{eq:ImpEqJ} then readily follows under the assumption that $\widetilde \Psi^p \neq 0$.

Special solutions to the Schr\"odinger equation [i.e., solutions not given by~\eqref{eq:PsiPsi} and \eqref{eq:ImpEqJ}] are obtained from considering the case $\widetilde \Psi^p = 0$, which implies that the components of the wave function $\Psi^i$ identically vanish. In this case, $\widetilde\Psi^p_q$ must satisfy
\be
\sum_q \widetilde \Psi^p_q=0 , \quad  \widetilde \Psi^p_0 =0. \label{app:spec-constraints}
\ee
The Schr\"odinger equation reduces to
\be
\left(\varepsilon_{p+q} - \frac{1}{2}J_K -\frac{1}{S}J\xi_q -E\right)\widetilde \Psi^p_q =0,
\ee
which implies that the energy is equal to
\be
E(p,q) = \varepsilon_{p+q} - \frac{1}{2}J_K -\frac{1}{S}J\xi_q.
\ee
For general values of momentum $p$ the constraints \eqref{app:spec-constraints} cannot the satisfied; the constraints can only be satisfied for $p=0$ or $p=\pi$ (the latter only when $N$ is even). This follows from the property $E(0,q)=E(0,-q)$ and $E(\pi,q)=E(\pi,-q)$. Consider the case $p=0$ as an example. For given $q$, the wave function components $\widetilde \Psi^0_q$ and $\widetilde \Psi^0_{-q}$ are nonzero and must satisfy
\be
\widetilde \Psi^0_q = - \widetilde \Psi^0_{-q}, \quad  |\widetilde \Psi^0_q|^2= |\widetilde \Psi^0_{-q}|^2=1/2. \label{app:wf_special}
\ee
The former follows from $\sum_q \widetilde \Psi^p_q=0$ and the latter from wave function normalization. The same reasoning applies to $q=\pi$. This then fully determines the structure of the special solutions to the Schr\"odinger equation.

How many of these special solutions are there? The answer depends on whether $N$ is odd or even. When $N$ is odd, there are $(N-1)/2$ pairs of momenta $q$ and $-q$ which yield a solution with energy $E(0,q)$. Instead, when $N$ is even, there are $N/2-1$ such solutions. This is fully consistent with the group theory analysis of Appendix~\ref{app:1D-group-theory}, since the special solutions constitute inversion-odd solutions, as may be seen from \eqref{app:wf_special}.

\section{Exact solution 1D chain when $J=0$ \label{app:1DJ0}}

This Appendix consider the special case $J=0$ of the 1D KH chain. When $J=0$ the problem simplifies and the structure of the solutions is slightly different. 

The wave function components are now related as
\be
\Tilde{\Psi}_q^p = -\frac{1}{\sqrt{2SN}}\frac{\varepsilon_p - \frac{1}{2}J_K-E}{\varepsilon_{p+q}-\frac{1}{2}J_K-E}\Tilde{\Psi},^p \label{app:PsiPsi}
\ee
and the implicit equation for the energy $E$ as a function of $p$ is given by
\be
\frac{\varepsilon_p+\frac{1}{2}J_K-E}{\varepsilon_p-\frac{1}{2}J_K-E}=\frac{1}{SN}\sum_q \frac{\frac12 J_K}{E -\varepsilon_{q}+\frac{1}{2}J_K}. \label{app:ImpEqJ}
\ee
These are trivial modifications of the general case. In contrast to the general case, however, now there are more special solutions to the Schr\"odinger equation which are not described by \eqref{app:PsiPsi} and \eqref{app:ImpEqJ}. We now construct these solutions. 

First, consider solutions for which $\Psi^i=0$. As before, this immediately implies that $\widetilde \Psi^p_0$ must satisfy
\be
\sum_q \widetilde \Psi^p_q=0 , \quad  \widetilde \Psi^p_0 =0.
\ee
The Schr\"odinger equation reduces to 
\be
\left(\varepsilon_{p+q} + \frac12J_K -E\right)\widetilde \Psi^p_q =0,
\ee
which implies that the energy is equal to
\be
E = \varepsilon_{p+q} + \frac12J_K .
\ee
Since $\varepsilon_{p}$ has the property $\varepsilon_{p+q} = \varepsilon_{-p-q}$, for given $p$ the wave function components $\widetilde \Psi^p_q$ and $\widetilde \Psi^p_{-2p-q}$ correspond to the same energy.

Consider now the structure of the wave function in more detail. It is easiest to first take $p=0$. Since $\varepsilon_{q}=\varepsilon_{-q}$ there are two values of $q$ for which $\widetilde \Psi^p_q$ is nonzero, namely $q$ and $-q$. Because of the constraint $\sum_q \widetilde \Psi^K_q=0$, these must satisfy
\be
\widetilde \Psi^0_q = - \widetilde \Psi^0_{-q}.
\ee
This means in particular that these solutions are odd under inversion. (Since $p=0$ is invariant under inversion, energy eigenstates are either even or odd under inversion.) A final constraint on the wave function is normalization, which in general reads
\be
|\widetilde \Psi^p| +\sum_q |\widetilde \Psi^p_q|^2=1,
\ee
and in this particular case reduces to
\be
 |\widetilde \Psi^0_q|^2= |\widetilde \Psi^0_{-q}|^2=1/2.
\ee

How many solutions of this type are there at $p=0$ for given $N$? The answer depends on whether $N$ is even or odd. If $N$ is odd, then there are $(N-1)/2$ of such pairs $q$ and $-q$ which yield a solution with energy $\varepsilon_{q} + \frac12JS$. Instead, if $N$ is even, then there are $N/2-1$ such solutions. Note further that when $N$ is even, $p=\pi$ is an allowed momentum, which is also inversion invariant. A similar argument then applies to $\widetilde \Psi^\pi_q$, yielding $N/2-1$ solutions with energy $\varepsilon_{\pi+q} + \frac12JS = -\varepsilon_{q} + \frac12JS$.

Now consider general momentum $p$. Only two components of the wave function corresponding to energy $\varepsilon_{p+q} + \frac12JS$ are nonzero: $\widetilde \Psi^p_q$ and $\widetilde \Psi^p_{-2p-q}$. These must satisfy the same constraint and normalization condition as before. Clearly these do not constitute eigenstates of inversion. How many solutions of this type are there? If $N$ is odd, then there are $(N-3)/2$ of such solutions and if $N$ is even, then there are $N/2-2$ such solutions. (There is one less solution compared to $p=0$ since there is one less admissible pair of momenta.)

We conclude by establishing a third and final class of special solutions. Inspection of the Fourier transformed Schr\"odinger equation suggests that solutions with energy $E= \varepsilon_{p} + \frac12JS$ may be found. Simply inserting this into the first equation gives
\be
-JS \widetilde \Psi^p +   \frac12 J\sqrt{2S} \Sigma = 0 \quad \rightarrow \quad \Psi^p =  \frac{1}{\sqrt{2S}} \widetilde\Sigma^p .
\ee
Inserting this then in the second equation simply gives 
\be
\left(\varepsilon_{p+q} -\varepsilon_{p} \right)\widetilde \Psi^p_q =0.
\ee
This equation must be solved subject to the constraints
\be
\Psi^p =  \frac{1}{\sqrt{2SN}}\sum_q \widetilde \Psi^p_q, \qquad \Psi^p = -\sqrt{2SN} \widetilde \Psi^p_0,
\ee
which may be combined into the condition
\be
0 = (2SN+1)\widetilde \Psi^p_0+\sum_{q\neq 0} \widetilde \Psi^p_q
\ee
No solutions exist for $p=0$ (similarly for $p=\pi$ when $N$ is even). For any other $p$ only two components of the wave function are nonzero: $\widetilde \Psi^p_0$ and $\widetilde \Psi^p_{-2p}$. These must then be related via
\be
\widetilde \Psi^p_{-2p} = (2SN+1)\widetilde \Psi^p_0.
\ee

\section{Spin-polaron in canonical spin wave expansion}\label{app:polaron}

In this appendix we describe in more detail how to compute the correlation function defined as
\be
C(r_i-r_j) =  N( \langle \bs^e_i \cdot \bS_j  \rangle - \langle \bs^e_i  \rangle \cdot  \langle \bS_j  \rangle ) /S, \label{app:C-definition}
\ee
with the spin wave expansion formalism. The ground state in which averages are calculated is defined in Eq.~\eqref{eq:polaronstate} and is thus given by
\be 
\ket{\Psi} =   \frac{1}{\sqrt{N}}\sum_{i = 1}^{N} d_{i}^{\dagger}\ket{\text{FM}} . \label{app:polaron-GS}
\ee 
For the operators we write
\be
\bs_i^e \cdot \bS_j = s_i^zS_j^z + \frac{1}{2}\left(s_i^+S_j^- + s_i^-S_j^+\right),
\ee
where 
\be
s_i^z = \frac{1}{2}(c_{i\up}^{\dagger}c_{i\up} - c_{i\down}^{\dagger}c_{i\down}), \quad s_i^+ = (s_i^-)^\dagger= c_{i\up}^{\dagger}c_{i\down},   
\ee
and the spin operators are expressed in terms of Holstein-Primakoff bosons as
\be
 S_j^z = S - a_j^{\dagger}a_j, \quad  S_j^+ = (S_j^-)^\dagger \simeq \sqrt{2S} a_j. 
\ee
The calculation of the averages then reduces to 
\begin{multline}
\left< \bs_i^e \cdot \bS_j\right> -  \left<s^{e,z}_i\right>\left<S^z_j\right> \simeq  \left< \bs_i^c \cdot \bS_j\right> -  S \left<s^{c,z}_i\right> \\
= - \left<s_i^za_j^{\dagger}a_j\right> + \sqrt{\frac{S}{2}} \left( \left<s_i^+a_j^\dagger  \right>+ \left< s_i^-a_j \right>\right), \label{app:averages}
\end{multline}
where we have neglected the contribution $\braket{ s^{e,z}_i} \braket{ a_j^{\dagger}a_j }$ since its contribution is of order $1/N^2$. 

The averages in \eqref{app:averages} are then computed in the state \eqref{app:polaron-GS} after using the expression \eqref{app:d-fermion} for the $d$-fermions. We compute the averages for three cases, i.e., $i-j=0,1,2$, which yields $C(r=0,1,2)$ given in Eq.~\eqref{eq:C(r)-SWT}.

\end{document}